\documentclass[10pt]{article}
\usepackage[dvips]{epsfig}
\usepackage[T1]{fontenc}
\usepackage[latin1]{inputenc}
\usepackage{graphicx}
\usepackage[english]{babel}
\usepackage{amsmath}
\usepackage{amssymb}
\usepackage{amsfonts}
\usepackage[T1]{fontenc}
\usepackage{color}
\usepackage{babel}
\usepackage{verbatim}
\usepackage[unicode=true,pdfusetitle,bookmarks=true,bookmarksnumbered=false,bookmarksopen=false,
 breaklinks=false,pdfborder={0 0 1},backref=false,colorlinks=true]{hyperref}
\hypersetup{linkcolor=blue,citecolor=blue}
\makeatletter
\usepackage[dvips]{epsfig}
\usepackage[T1]{fontenc}
\usepackage{color}
\usepackage{pdflscape}
\begin{document}

\title{\bf   New Classes of Spherically Symmetric,  Inhomogeneous Cosmological
Models}
\author{Metin G{\" u}rses\thanks{%
email: gurses@fen.bilkent.edu.tr}~~and~~Yaghoub Heydarzade\thanks{%
email: yheydarzade@bilkent.edu.tr}
\\{\small Department of Mathematics, Faculty of Sciences, Bilkent University, 06800 Ankara, Turkey}}

\maketitle

\begin{abstract}
We present two classes of inhomogeneous, spherically symmetric solutions of the Einstein-Maxwell-Perfect Fluid field equations with cosmological constant generalizing  the Vaidya-Shah solution. Some special limits of our solution reduce to the known inhomogeneous charged perfect fluid solutions of the Einstein field equations and under some other limits we obtain new  charged and uncharged solutions with cosmological constant. Uncharged solutions in particular represent cosmological models where the universe may undergo a topology change and in between is a mixture of two different Friedmann-Robertson-Walker universes with  different spatial curvatures. We show that there exist some spacelike surfaces  where the Ricci scalar and pressure of the fluid diverge  but the mass density of the fluid distribution remains finite. Such spacelike surfaces are known as (sudden) cosmological singularities.    We study the behavior of our new solutions in their general form as the radial distance goes to zero and infinity. Finally, we briefly address the null geodesics and apparent horizons associated to the obtained solutions.
\end{abstract}

\section*{I.~Introduction}
In the last two decades there is an increasing interest in studying and finding exact inhomogeneous cosmological solutions in general relativity. Observational effects of inhomogeneity in cosmology are discussed in several works. Among these we note that the collection of articles in \cite{ellis}-\cite{bolejku} are worth mentioning.

There are many reasons to study inhomogeneous cosmological models in general relativity. Among these, the following three mentioned by Ellis \cite{ellis} (see also the references therein) are
important. Local inhomogeneity may effect the averaged large scale dynamics of the universe (see also \cite{anderson} and the references therein
), local inhomogeneity may effect the photon propagation hence may change the cosmological observations and the inhomogeneity at Hubble scale with the violation of Copernicus principle may lead to acceleration of the universe (see also \cite{marra} and the references therein). In his book \cite{kr0} Krasinski gives other reasons such as the formation of voids and interaction of the cosmic microwave background radiation with matter in the universe can be explained by  exact solutions of the Einstein field equations in an  inhomogeneous spacetime. For all these reasons it worths finding new inhomogeneous solutions of Einstein's field equations.

Spherically symmetric cosmological models were studied previously by many authors \cite{bolejku}-\cite{thak}.
Historically, Lemaitre \cite{lem, kr} and McVittie metrics \cite{mcv} can be considered as the first inhomogeneous solutions of the Einstein-perfect fluid field equations. Recently, it has been shown that the McVittie solution represents a black hole in an expanding universe \cite{kaloper, lake}. Charged version of the McVittie solution is known as the Vaidya-Shah metric \cite{vaidya1}-\cite{vaidya3} which is a spherically symmetric solution of the Einstein-Maxwell-perfect fluid field equations. This metric
of this solution is given as follows
\begin{equation}
ds^2=-A^2\, dt^2+B^2\,(dr^2+r^2\, d\theta^2+r^2\, \sin^2(\theta)\, d\phi^2), \label{vs1}
\end{equation}
where
\begin{eqnarray}
&&A=\frac{[1-(M^2-Q^2) \frac{1+k r^2}{4a^2(t)\,r^2}]}{\left[1+M \frac{\sqrt{1+k r^2}}{a(t)r}+(M^2-Q^2) \frac{1+k r^2}{4a^2(t)\,r^2}\right]},\\
&&B=\frac{a(t)}{1+k r^2}\, \left[ 1+M \frac{\sqrt{1+k r^2}}{a(t)r}+(M^2-Q^2) \frac{1+k r^2}{4a^2(t)\,r^2} \right] \label{bb122},
\end{eqnarray}
where $a(t)$ is any arbitrary function of time $t$, $M$ and  $Q$ are constants representing the conserved quantities of mass and charge, and $k$ is also a constant.

Pressure, mass and charge densities are respectively given by

\begin{eqnarray}
&&8 \pi p=-\frac{2}{A}\, \left(\frac{\ddot{a}(t)}{a(t)}-\frac{\dot{a}^2(t)}{a^2(t)} \right)-3 \frac{\dot{a}^2(t)}{a^2(t)}- \frac{4ka(t)}{A B^3  (1+k r^2)^3},\label{vsp}  \label{vs2}\\
&&8 \pi \rho=3 \,\frac{\dot{a}^2(t)}{a^2(t)}+ \frac{6k}{ a^2(t)}\, \left[1+M \frac{\sqrt{1+k r^2}}{a(t)r}+(M^2-Q^2) \frac{1+k r^2}{4a^2(t)\,r^2}\right]^{-3} \left[2 +M \frac{\sqrt{1+k r^2}}{a(t)r} \right]\label{vsrho},\\
&&4 \pi \sigma=-\frac{3 k Q}{ a^3(t)}\, \frac{\sqrt{1+ kr^2}}{r}\,\left[1+M \frac{\sqrt{1+k r^2}}{a(t)r}+(M^2-Q^2) \frac{1+k r^2}{4a^2(t)\,r^2}\right]^{-3}.   \label{vssigma}
\end{eqnarray}

The uncharged ($Q=0$)  Vaidya-Shah solution is more general than the McVittie solution. McVittie solution corresponds to $Q=0$ and $k=0$. In spite of this fact, Vaidya-Shah solution is sometimes named as the charged McVittie solution. The Vaidya-Shah metric reduces to the Reissner-Nordstr{\" o}m metric when $k=0$ and $a(t)=1$  in isotropic coordinates. Note that the charge density (\ref{vssigma}) for Vaidya-Shah solution vanishes as $k=0$ but the Maxwell's field $F_{01}$ remains nonzero. Vaidya-Shah solution \cite{vaidya3}  has been studied by several authors \cite{feroni1}-\cite{dasilva}, \cite{lake}, and like the McVittie solution it has been shown that it describes a charged black hole in an expanding universe. The charged and uncharged cosmological black holes were also discussed in the works \cite{kr1}-\cite{kr5}.

In this work, we start with the spherically symmetric metric in the isotropic coordinates in four dimensions
\begin{equation}
ds^2=-a^2 dt^2+b^2\left(dr^2+r^2\,d\theta^2+r^2\, \sin^2(\theta)\, d\phi^2\right),
\end{equation}
where $a$ and $b$ are differentiable functions of $t$ and $r$. We first show that the Einstein Maxwell-Perfect Fluid field equations with cosmological constant reduce to a single nonlinear ordinary differential equation for the function $b(t,r)$ (Theorem 1). Then we solve this differential equation as general as possible. We use the method of separation of variables and find two distinct classes of solutions (Theorem 2). For the charged case, we have the following distinct solutions:

\vspace{0.5cm}
\noindent
{\bf Class 1}:

\begin{equation}
b(t,r)=\frac{\delta}{\sqrt{c_{0}+c_{1}\,r^2}\,\sqrt{c_{2}+c_{3}\,r^2}}+\beta(t)\, \frac{1}{c_{0}+c_{1}\, r^2}+\frac{\gamma}{\beta(t)}\,\frac{1}{c_{2}+c_{3}\, r^2},
\end{equation}

\noindent
and

\vspace{0.5cm}
\noindent
{\bf class 2}:

\begin{equation}
b(t,r)=\nu_{0}(r)+\frac{a(t)}{ b_{0}+b_{1}\, r^2},
\end{equation}
where $a(t)$ and $\beta(t)$ are arbitrary functions of $t$, $\nu_{0}(r)$ is an arbitrary function of $r$ while $b_{0}, b_{1}, c_{0}, c_{1}, c_{2}, c_{3}, \delta $ and $\gamma$ are arbitrary constants. For the uncharged case, the above solutions reduce to the following distinct solutions:

\vspace{0.5cm}
\noindent
{\bf Class 1}:

\begin{equation}
b(t,r)=\left(\frac{\delta}{2\, \sqrt{\beta(t)}}\, \frac{1}{\sqrt{c_{2}+c_{3}\,r^2}}
+\frac{\sqrt{\beta(t)}}{\sqrt{c_{0}+c_{1}\,r^2}} \right)^2,
\end{equation}

\noindent
and

\vspace{0.5cm}
\noindent
{\bf Class 2}:
\begin{equation}
b(t,r)=\frac{b_{2}+a(t)}{ b_{0}+b_{1}\, r^2},
\end{equation}
where $b_{2}$ is  also an arbitrary constant. For all the above cases, we found  $a(t,r)=q(t)\, \frac{\dot{b}}{b}$. We show that, in particular for the uncharged case, the
first class of solutions exhibit a cosmological model describing a universe as  a mixture of two different Friedmann-Robertson-Walker universes with different spatial curvatures. If the signs of the spatial curvatures are different then we show that there is a possibility of the change of topology of the universe. If the spatial curvatures turns out to be the same the spacetime becomes a single FRW universe. We then study the asymptotical properties of our solutions. We show that the six parameter solution which is the generalization of the  Vaidya-Shah solution (\ref{vs1})-(\ref{vssigma}) is nonsingular as the radial distance goes to zero and to infinity (Theorem 3). The uncharged limit ($Q=0)$ of our solutions generalize the McVittie solution. We show that there are surfaces  $\Sigma_{1}$ ($b(t,r)=0$) and $\Sigma_{2}$ ($a(t,r)=0$) where the Ricci scalar diverges (spacetime singularities). $\Sigma_{1}$ is a timelike but $\Sigma_{2}$ is a spacelike surface. Physical constraints eliminate the timelike surfaces $\Sigma_{1}$ and there  remain only the spacelike singular surfaces $\Sigma_{2}$.  This surface is commonly named as the cosmological singularity \cite{penrose} where the mass density is regular but the pressure diverges on this surface. This surface is  also  called a ``sudden cosmological singularity'' \cite{barrow1}-\cite{lake2004}. We also obtain the apparent horizons of our solutions which correspond to null (constant) areal  distance surfaces. We give a plot of null geodesics, apparent horizons and singular surface $\Sigma_{2}$ for the $N=2$ uncharged solution for particular values of the parameters of the solution.

The layout of the paper is as follows. In Section II, we simplify and reduce the field equations into a single ordinary nonlinear  differential equation. In Section III, we solve the resulting differential equation by the use of the method of
separation of variables and obtain two different distinct solutions. In Section IV, we obtain the asymptotic behaviours of our solutions and show that the corresponding spacetimes are nonsingular with respect to the asymptotic values of $r$. In Section V, we study all possible special limits of our solutions. In Section VI, we study the uncharged versions of our solutions. In Section VII, we investigate the possible apparent horizons and null geodesics of the charged and uncharged solutions. In Appendix A, we write the differential equation obtained in Section II in a different form and in Appendixes B-F  we give the long expressions obtained in Sections IV and V. In the last Appendix we give the mass densities when $a \to 0$ and $a \to \infty$ respectively.

\section*{II.~Field Equations of the Charged Fluids in Four Dimensions}
We consider the Einstein-(anti) de Sitter-Maxwell-Perfect Fluid field equations
\begin{eqnarray}\label{111a}
&&G_{\mu \nu}+\Lambda g_{\mu\nu}=8\pi T_{\mu\nu} +E_{\mu\nu} \label{eqns1},
\end{eqnarray}
where
\begin{eqnarray}
&&T_{\mu\nu}=(p+\rho) u_{\mu} u_{\nu}+p\,g_{\mu \nu},\label{1dd1a}\\
&&  E_{\mu\nu}=2 \left( F_{\mu \alpha}\, F_{\nu}\,^{\alpha}-\frac{1}{4} F_{\alpha \beta}F^{\alpha \beta}\,g_{\mu \nu} \right), \label{vij}\\
&&\nabla_{\alpha}\,F^{\mu \alpha}=4 \pi \sigma u^{\mu},\label{namana}
\end{eqnarray}
where  $\Lambda$, $T_{\mu\nu}$, $E_{\mu\nu}$,  and $F_{\mu\nu}$ are the cosmological constant, energy-momentum
tensor of the perfect fluid, Maxwell and Faraday tensors, respectively.
To obtain our solutions, we consider the spherical symmetric metric
\begin{equation}\label{1c}
ds^2=-a^2 dt^2+b^2\left(dr^2+r^2\,d\theta^2+r^2\, \sin^2(\theta)\, d\phi^2\right),
\end{equation}
where $a$ and $b$ are generic functions of both the time $t$ and radial coordinate $r$, i.e $a=a(t,r)$ and $b=b(t,r)$. Regarding the spherical symmetry
in the spacetime metric (\ref{1c}), the only non-vanishing component of the antisymmetric  electromagnetic Faraday tensor is
\begin{equation}\label{2d}
F_{01}=\psi,
\end{equation}
where $\psi=\psi(t,r)$. Using the non-zero source Maxwell equation (\ref{namana})
and the metric (\ref{1c}), we obtain
\begin{equation}\label{psi1}
\dot \psi=\psi\left(\frac{\dot a}{a}-\frac{\dot b}{b}  \right),
\end{equation}
and
\begin{equation}\label{sigma1}
4\pi \sigma=\frac{1}{ab^2}\left(\psi^{\prime}+\psi\left(\frac{b^\prime}{b}-\frac{a^\prime}{a}  \right) +\frac{2}{r}\psi  \right),
\end{equation}
where $\sigma=\sigma(t,r)$ is the charge density and the dot and prime signs denote the derivatives with respect to time
and radial coordinates, respectively.

On the other hand, using  (\ref{1dd1a}), (\ref{vij}) and (\ref{2d})
and considering the perfect fluid velocity vector as $u_{\mu}=a \, \delta^{0}_{\mu}$, the $00$ component of
the Einstein-Maxwell-Perfect Fluid equations (\ref{111a}) gives
\begin{equation}\label{rho1}
8\pi \rho(t,r)=-\frac{1}{a^2 b^2}\psi^2 -\frac{4bb^{\prime} a^{2}-{b^{\prime}}^2 r a^2 -3b^2r\dot
b^2 +2brb^{\prime\prime}a^2}{a^2 b^4 r}-\Lambda.
\end{equation}
The $01$ components reads as
\begin{equation}\label{int}
\dot b b^{\prime} a-b \dot b^{\prime}a+b\dot ba^{\prime}=0,
\end{equation}
while the $11$ and $22$ (or $33$) components lead to
\begin{equation}\label{p1}
8\pi p(t,r)=\frac{1}{a^2 b^2}\psi^2 +\frac{1}{a^3 b^4 r}\left(-2b^3 r \ddot ba +2b^3 r \dot a \dot b + 2bra^2 a^{\prime}b^{\prime} +{b^{\prime}}^2 r a^3 +2bb^{\prime} a^3 -ab^2
r\dot b^2 +2b^2 a^2 a^{\prime}   \right)+\Lambda,
\end{equation}
and
 \begin{eqnarray}\label{p2}
8\pi p(t,r)&=&-\frac{1}{a^2 b^2}\psi^2 \nonumber\\
&&+\frac{1}{a^3 b^4 r}\left(-2b^3 r \ddot b
a +2b^3 r \dot a \dot b + b^{2}a^2 a^{\prime} + bb^{\prime}a^3 +brb^{\prime\prime} a^3 -ab^2
r\dot b^2 -{b^{\prime}}^2 r a^{3}+b^2 r a^2 a^{\prime\prime}\right)+\Lambda,\nonumber\\
\end{eqnarray}
 respectively. We can integrate the equation (\ref{int}) to obtain
\begin{equation}\label{a1}
a(t,r)=q\frac{\dot b}{b},
\end{equation}
where $q=q(t)$ and $\dot b(t,r)\neq0$. One notes that for $\dot b(t,r)=0$, the equation
(\ref{int}) disappears. Using (\ref{psi1}), we  arrive at
\begin{equation}\label{psi2}
\psi(t,r)=hq \frac{\dot b}{b^2},
\end{equation}
where $h=h(r)$ is an arbitrary functions of $r$. Using (\ref{a1}) and (\ref{psi2}), the charge density $\sigma$ in (\ref{sigma1})
takes the following form
\begin{equation}\label{charge}
 4\pi \sigma(t,r)=\frac{1}{rb^3}\left(r h^{\prime}+2h \right).
\end{equation}
Then, the total charge  $Q_{T}$  in a spherical region with radius $R_{0}$ ($t$ constant, $r$ constant regions) can be obtained as

\begin{equation}
Q_{T}=\iiint\sigma dV=4 \pi \int_{o}^{R_{0}}\, (r^2 h^{\prime}+2 r h) dr.
\end{equation}
Hence, the total charge in this volume is given by
\begin{equation}\label{qt}
Q_T =r^2 h\mid_{r=R_0}.
\end{equation}

Finally, the equations (\ref{p1}) and (\ref{p2}) reduce to
\begin{eqnarray}\label{denk2}
r b^2 {\dot b}^{\prime\prime}=4 r b b^{\prime} \dot b^{\prime}+b^2 \dot b^{\prime}-2 r \dot b {b^{\prime}}^2+2 r h^2 \dot b.
\end{eqnarray}
One can integrate the differential equation (\ref{denk2}) with respect time and obtain the following  second order   ordinary nonlinear differential equation for $b$

\begin{equation}\label{denk3}
-r b b^{\prime\prime}+2 r {b^{\prime}}^2+b b^{\prime}-2 r h^2+ h_{1} b=0,
\end{equation}
where $h_1=h_{1}(r)$ is a new arbitrary function of $r$.

\vspace{0.4cm}
To summarize what we have till now, we introduce the following  theorem.
\vspace{0.4cm}
\\{\textbf{Theorem 1}}:
\textit{ Einstein field equations of a charged perfect fluid with a cosmological constant of  a spherically symmetric spacetime reduce to the following subclasses.
 \begin{description}
 \item[(i)] For $\dot b(t,r)\neq0$, the field equations reduce to a single ordinary nonlinear differential
 equation, Eq (\ref{denk3}),  with two arbitrary functions of r, $h$ and $h_1$ functions.  Then, the metric function $a(t,r)$ and the charge density $\sigma(t,r)$
are given by (\ref{a1}) and (\ref{charge}) respectively, and  the energy density $\rho(t,r)$ in (\ref{rho1}) and the pressure $p(t,r)$ in (\ref{p1}) (or (\ref{p2}))  respectively read as
\begin{equation}\label{rhotho}
 8 \pi\rho(t,r)=\frac{3}{q^2}+\frac{3 h^2}{ b^4}-\frac{1}{ rb^{4}}\left(3 r  {b^{\prime}}^2+6  bb^{\prime}+2 h_{1}b \right)-\Lambda,
\end{equation}
and
\begin{equation}\label{ptho}
8 \pi p(t,r)=-\frac{3}{q^2}+\frac{ h^2}{ b^4}+\frac{1}{ rq^3 b{^4} \dot
b}\, \left(2 b q^3\,(r b^{\prime}+b)\, {\dot b}^{\prime} - r q^3\,
{\dot b}\, {b^{\prime}}^2
 +2rb^5 \dot q\right)+\Lambda.
\end{equation}
\item[(ii)] For $\dot b(t,r)=0$, there is no $01$ component for the field equations,
then the equation (\ref{int}) and the relation between the metric functions
as (\ref{a1}) disappears.
For this case, the Maxwell equation (\ref{namana}) gives $\psi(t,r)=h(r)a(t,r)$, and the equations (\ref{rho1}),  (\ref{p1}) reduce to
\begin{eqnarray}
&&8\pi \rho(t,r)=-\frac{1}{a^2 b^2}\psi^2 -\frac{4bb^{\prime} -{b^{\prime}}^2 r +2brb^{\prime\prime}}{b^4 r}-\Lambda,\\
&&8\pi p(t,r)=\frac{1}{a^2 b^2}\psi^2 +\frac{1}{a^3 b^4 r}\left( 2br a^{\prime}b^{\prime} +{b^{\prime}}^2 r a +2bb^{\prime} a +2b^2 a^{\prime}   \right)+\Lambda,\label{ngk}
\end{eqnarray}
where $b(r)$ should satisfy the following equation
 \begin{equation}
2rab^{2}h^{2}+ 2rb a^{\prime}b^{\prime} +2ra{b^{\prime}}^2 +abb^{\prime}  + a^{\prime}b^2-rabb^{\prime\prime} -ra^{\prime\prime}b^2 =0.
\end{equation}
Solving this single differential equation with three unknown functions
$h(r)$, $a(t,r)$ and $b(r)$ is not possible except by supposing relations
between these functions. One possible ansatz can be considering a specific
equation of state for the perfect fluid, leading to a relation between $a(t,r)$ and $b(r)$ functions.
\end{description}}

In this work, we consider only the general dynamical case, i.e $\dot b(t,r)\neq 0$, and then
our aim is to solve the nonlinear ordinary differential equation (\ref{denk3})
for the metric function $b(t,r)$. In the next sections, we will solve this
equation and determine all of our unknown functions $a(t,r)$, $\psi(t,r)$,
$\sigma(t,r)$, $\rho(t,r)$ and $p(t,r)$ accordingly.

\section*{III.~Exact Solutions of the Field Equations}
The main aim of this section is to find solutions of the equation (\ref{denk3}). For this purpose, we use the method of separation of variables. Although the equation (\ref{denk3}) is a nonlinear ordinary differential equation, we can use this method by equating the coefficients of the products of the time dependent functions to zero. Let
\begin{equation}\label{bb}
b(t,r)=\sum_{n=0}^{N}\,\alpha_{n}(r)\, \beta_{n}(t),
\end{equation}
where $\alpha_{n}(r)$ and $\beta_{n}(t)$ are  all independent functions of $r$ and $t$, respectively such that $n=0, 1, 2, \cdots, N$.
 There are $N+1$ number of functions $\alpha_{n}(r)$ depending on $r$ in (\ref{bb}) and 2 arbitrary functions $h_{1}(r)$ and $h(r)$ in the main equation (\ref{denk3}). Then, totally we have $N+3$ functions of $r$. The functions $\beta_{n}(t)$ ($n=0,1,2,\cdots, N$) are left arbitrary but independent functions of $t$. The time independent term $2 r h^2$
 in the main equation (\ref{denk3}) forces us to choose one of the time dependent functions $\beta_{n}(t)$ ($n=0,1,2,\cdots, N$) to be a constant. Thus, without losing any generality, we let $\beta_{0}=1$. Hence, we have
\begin{equation}\label{bb1}
b(t,r)=\alpha_{0}(r)+\sum_{n=1}^{N}\,\alpha_{n}(r)\, \beta_{n}(t).
\end{equation}
 By inserting (\ref{bb1}) in (\ref{denk3}), we obtain more than $2N+1$ equations. This means that when $N >2$, the number of equations becomes more than the number of unknown functions (an overdetermined system). Hence, we use the ansatz (\ref{bb}) only for $N=2$ and for $N=1$, and we  investigate these cases in detail in the
sections III A and III B. Before to proceed, we refer the reader
to   Appendix A summarizing the method introduced in  \cite{lie, exact} for solving the equation (\ref{denk3})
for the uncharged case, where some particular solutions are also addressed. To produce the most generic solutions including the
charge, the approach in \cite{lie, exact} seems not suitable for us and we will follow  the method of separation of variables as discussed above.

\subsection*{A.~Solutions for $N=2$}\label{n2sol}
 Letting $N=2$,  we have
\begin{equation}\label{4d}
b(t,r)=\alpha_{0}(r)+\beta_{1}(t)\alpha_{1}(r)
+\beta_{2}(t)\alpha_{2}(r),
\end{equation}
where as mentioned before, $\alpha_{0}$, $\alpha_{1}$ and $\alpha_{2}$ are functions of $r$ and $\beta_{1}$ and $\beta_{2}$ are functions of $t$.  In Eq.(\ref{denk3}), when the function $b$ in (\ref{4d}) is inserted, the coefficients of the time dependent functions $\beta_{1}^2$, $\beta_{2}^2$, $\beta_{1}$ and $\beta_{2}$ are set to zero and functions $\alpha_{0}$, $\alpha_{1}$ and $\alpha_{2}$ satisfy the following equations
\begin{eqnarray}
&& \beta_{1}^2:~~-r \alpha_{1} \alpha_{1}^{\prime \prime}+2 r (\alpha_{1}^{\prime})^2+\alpha_{1} \alpha_{1}^{\prime}=0, \label{denk6}\\
&& \beta_{2}^2:~~-r \alpha_{2} \alpha_{2}^{\prime \prime}+2 r (\alpha_{2}^{\prime})^2+\alpha_{2} \alpha_{2}^{\prime}=0, \label{denk7} \\
&& \beta_{1}:~~-r \alpha_{0} \alpha_{1}^{\prime \prime}-r \alpha_{1} \alpha_{0}^{\prime \prime}+4 r \alpha_{0}^{\prime} \alpha_{1}^{\prime}+\alpha_{0} \alpha_{1}^{\prime}+ \alpha_{1} \alpha_{0}^{\prime}+h_{1} \alpha_{1}=0, \label{denk8} \\
&&\beta_{2}:~~-r \alpha_{0} \alpha_{2}^{\prime \prime}-r \alpha_{2} \alpha_{0}^{\prime \prime}+4 r \alpha_{0}^{\prime} \alpha_{2}^{\prime}
+\alpha_{0} \alpha_{2}^{\prime}+ \alpha_{2} \alpha_{0}^{\prime}+h_{1} \alpha_{2}=0. \label{denk9}
\end{eqnarray}
The remaining equation depends how the functions $\beta_{1}$ and $\beta_{2}$ are related which reads as
\begin{eqnarray}\label{eqn4}
&&-r \alpha_{0} \alpha_{0}^{\prime \prime}+2 r (\alpha_{0}^{\prime})^2+\alpha_{0} \alpha_{0}^{\prime}-\kappa r h^2 +h_{1}\, \alpha_{0}\nonumber \\
&&+\beta_{1}\, \beta_{2}\left(-r \alpha_{1} \alpha_{2}^{\prime \prime}-r \alpha_{2} \alpha_{1}^{\prime \prime}+4 r \alpha_{1}^{\prime}\,\alpha_{2}^{\prime}+\alpha_{1}\, \alpha_{2}^{\prime}+\alpha_{2}\, \alpha_{1}^{\prime}\right)=0.
\end{eqnarray}
General solutions of (\ref{denk6}) and (\ref{denk7}) are given by
\begin{equation}\label{gfg}
\alpha_{1}=\frac{1}{c_{0}+c_{1}\, r^2},~~~\alpha_{2}=
\frac{1}{c_{2}+c_{3}\, r^2},
\end{equation}
where $c_{0}$, $c_{1}$, $c_{2}$ and $c_{3}$ are arbitrary constants. In (\ref{denk8}) and (\ref{denk9}), the function $\alpha_{0}$ satisfies a second order linear differential equation.
Multiplying (\ref{denk8}) by $\alpha_{2}$ and (\ref{denk9}) by $\alpha_{1}$, and subtracting them, we obtain $\alpha_{0}$ as
\begin{equation}
\alpha_{0}^4=c_{4}\,\frac{1}{r} (\alpha_{2}\,\alpha_{1}^{\prime}-
\alpha_{1}\,\alpha_{2}^{\prime}) =\frac{2c_4\left(c_3c_0-c_1 c_2  \right)}
{\left(c_{0}+c_{1}\,r^2\right)^2 \left(c_{2}+c_{3}\,r^2\right)^2},
\end{equation}
or
\begin{equation}\label{ggg}
\alpha_{0}=\frac{\delta}{\sqrt{c_{0}+c_{1}\,r^2}\,\sqrt{c_{2}+c_{3}\,r^2}},~~\delta=\pm\sqrt[4]{2 c_{4}\,(c_{3}\,c_{0}-c_{1}\,c_{2})},
\end{equation}
where $c_4$ is an arbitrary constant, and we have the condition $c_4(c_{3}\,c_{0}-c_{1}\,c_{2})>0$.
As we will see in the classification of the possible solutions, the negative sign
of $\delta$ is not physical due to its identification relation to mass.  The equation (\ref{denk9}) can be considered as the definition of the function $h_{1}$. Hence, we have solved all equations (\ref{denk6})-(\ref{denk9}). There remains only $h$ function to be determined. For
determining function $h$, there are two possibilities as follows.
\begin{description}
\item[(i)]  \textbf{If $\beta_{1}$ and $\beta_{2}$ have no relations (if $\beta_1 \beta_2 \neq constant$). }
\\
For this case,
using (\ref{eqn4}), we have
\begin{eqnarray}
&&-r \alpha_{0} \alpha_{0}^{\prime \prime}+2 r (\alpha_{0}^{\prime})^2+\alpha_{0} \alpha_{0}^{\prime}-2 r h^2+h_{1}\alpha_0=0, \label{denk11}\\
&&-r \alpha_{1} \alpha_{2}^{\prime \prime}-r \alpha_{2} \alpha_{1}^{\prime \prime}+4 r \alpha_{1}^{\prime}\,\alpha_{2}^{\prime}+\alpha_{1}\, \alpha_{2}^{\prime}+\alpha_{2}\, \alpha_{1}^{\prime}=0. \label{denk12}
\end{eqnarray}
Now, the equation (\ref{denk11}) can be considered as the definition of the function $h$ but the last equation (\ref{denk12}) gives $c_{1}=c_{3}=0$ which means that the function $b$ depends only on $t$ which is not our desired solution in general.
\item[(ii)]  \textbf{If $\beta_{1}\beta_{2}=\gamma$ where $\gamma$ is a constant.}\\ For this case, we have the following single differential equation
\begin{eqnarray}\label{eqn5}
&&-r \alpha_{0} \alpha_{0}^{\prime \prime}+2 r (\alpha_{0}^{\prime})^2+\alpha_{0} \alpha_{0}^{\prime}-2 r h^2 +h_{1}\, \alpha_{0}\nonumber \\
&&+\gamma\,\left(-r \alpha_{1} \alpha_{2}^{\prime \prime}-r \alpha_{2} \alpha_{1}^{\prime \prime}+4 r \alpha_{1}^{\prime}\,\alpha_{2}^{\prime}+\alpha_{1}\, \alpha_{2}^{\prime}+\alpha_{2}\, \alpha_{1}^{\prime}\right)=0.
\end{eqnarray}
This equation can be considered as the definition of the function $h$. Thus,
by this consideration,  we can solve the equation (\ref{denk3}) completely.
Then, the function $b(t,r)$ takes the form of
\begin{equation}\label{bb3}
b(t,r)=\frac{\delta}{\sqrt{c_{0}+c_{1}\,r^2}\,\sqrt{c_{2}+c_{3}\,r^2}}+\beta(t)\, \frac{1}{c_{0}+c_{1}\, r^2}+\frac{\gamma}{\beta(t)}\,\frac{1}{c_{2}+c_{3}\, r^2}.
\end{equation}
\end{description}
\vspace{0.5cm}
\subsection*{B.~ Solutions for $N=1$\label{CCC}}
Considering $N=1$, we have
\begin{equation}\label{bigul}
b(t,r)=\nu_{0}(r)+\beta(t)\, \nu_{1}(r),
\end{equation}
where here $\nu_{0}(r)$ and $\nu_{1}(r)$ are functions of $r$ and $\beta(t)$ is a function of $t$. By inserting the function $b(t,r)$ in (\ref{bigul}) in the equation (\ref{denk3}),  $\nu_{0}(r)$ and $\nu_{1}(r)$ should satisfy the following equations
\begin{eqnarray}
&&\beta^2:~~-r \nu_{1} \nu_{1}^{\prime \prime}+2 r (\nu_{1}^{\prime})^2+\nu_{1} \nu_{1}^{\prime}=0, \label{denk13}\\
&&\beta:~~-r \nu_{0} \nu_{1}^{\prime \prime}-r \nu_{1} \nu_{0}^{\prime \prime}+4 r \nu_{0}^{\prime} \nu_{1}^{\prime}
+\nu_{0} \nu_{1}^{\prime}+ \nu_{1} \nu_{0}^{\prime}+h_{1} \nu_{1}=0, \label{denk14} \\
&&\beta^{0}:~~-r \nu_{0} \nu_{0}^{\prime \prime}+2 r (\nu_{0}^{\prime})^2+\nu_{0} \nu_{0}^{\prime}-2 r h^2+\nu_{0}\,h_{1}=0. \label{denk15}
\end{eqnarray}
General solution of (\ref{denk13}) is
\begin{equation}
\nu_{1}=\frac{1}{b_{0}+b_{1}\, r^2},
\end{equation}
where $b_{0}$ and $b_{1}$ are arbitrary constants.
Equations (\ref{denk14}) and (\ref{denk15}) can be considered as the definitions of the functions $h$ and $h_1$. Hence, $\nu_{0}(r)$
function is left arbitrary. Then,  $b(t,r)$ takes the following form
\begin{equation}\label{n1bb}
b(t,r)=\nu_{0}(r)+\frac{\beta(t)}{ b_{0}+b_{1}\, r^2}.
\end{equation}
\\
\vspace{0.1cm}
Thus, the following theorem represents the summary of what is done till now.
\vspace{0.1cm}
\\{\textbf{Theorem 2}}:
\textit{ The most general solutions of the ordinary nonlinear differential
equation (\ref{denk3}) by the method of separation of variables are  given in two classes: The first one containing one arbitrary function of $t$ and six arbitrary parameters is
\begin{equation}\label{qltc}
b(t,r)=\frac{\delta}{\sqrt{c_{0}+c_{1}\,r^2}\,\sqrt{c_{2}+c_{3}\,r^2}}+\beta(t)\, \frac{1}{c_{0}+c_{1}\, r^2}+\frac{\gamma}{\beta(t)}\,\frac{1}{c_{2}+c_{3}\, r^2},
\end{equation}
corresponding to $N=2$
and the second one containing two arbitrary constants and two arbitrary functions where one depends on $r$ and the other depends on $t$
 \begin{equation}
b(t,r)=\nu_{0}(r)+\frac{\beta(t)}{ b_{0}+b_{1}\, r^2},
\end{equation}
corresponding to $N=1$.}\\
A different approach is given in Appendix I for solving (\ref{denk3}). Such an approach was introduced in \cite{lie} for the uncharged case (see also \cite{bolejku, exact}).

%
\section*{IV.~Properties of the Solutions to the Field Equations}
In this section, we first investigate singular structure of the obtained
spacetimes. There are surfaces $\Sigma_{1}$ and $ \Sigma_{2}$  where the pressure $p$ and mass density $\rho$ diverge. Then, we explicitly check the properties of the general solutions
for both the cases of $N=2$ and $N=1$  as $r$ goes to zero and tends to infinity, in detail. Furthermore, we will address
some specific subclasses of these general solutions and study their properties also in the next sections.
\subsection*{A.~Singular Structure of the Solutions}
Here, we assume that $c_{0}, c_{1}, c_{2}$ and $c_{3}$ are non negative constants. Regarding the field equations (\ref{eqns1}), the scalar curvature (Ricci
scalar) is given by $R=8\pi(\rho-3 p)+4\Lambda$. Hence, if any one of the quantities $p$ or $\rho$ is singular on some surfaces then they are the spacetime singularities. Regarding (\ref{a1}), (\ref{rhotho}) and (\ref{ptho}), if the functions $b$ and $a$ vanish on some surfaces then either $p$ or $\rho$ diverges. Hence, we will focus on the surfaces $\Sigma_{1}=\{(t,r)\in U | b(t,r)=0 \}$ and $\Sigma_{2}=\{(t,r) \in U | a(t,r)=0\}$. Here, $U$ is a part of spacetime where $-\infty<t<\infty,~r \ge 0$. In the following, we will explore  these singular surfaces.

\vspace{0.5cm}
\noindent
{\bf 1. Singular Surfaces for  the Class of $N=2$}\\
\begin{description}
\item[(i)] \textbf{Surface $\Sigma_{1}$}

Letting $X=\left(\frac{c_{0}+c_{1} r^2}{c_{2}+c_{3} r^2}\right)^{\frac{1}{2}}$, then $b(t,r)=0$ leads to the following equation
\begin{equation}
\frac{\gamma}{\beta} X^2+\delta X+\beta=0.
\end{equation}
When $\gamma \ne 0$ this equation has real solutions only when $\delta^2-4 \gamma \ge 0$. Then, there are two different dynamical surfaces given by (depending on the sign of $\delta$)
\begin{equation}
\left(\frac{c_{0}+c_{1} r^2}{c_{2}+c_{3} r^2}\right)^{\frac{1}{2}}=\frac{-\delta \pm \sqrt{\delta^2-4 \gamma}}{2 \gamma}\, \beta(t).
\end{equation}
When $\gamma=0$ we have
\begin{equation}
\left(\frac{c_{0}+c_{1} r^2}{c_{2}+c_{3} r^2}\right)^{\frac{1}{2}}=-\frac{\beta(t)}{\delta}.
\end{equation}
The normal vectors of these surfaces satisfy
\begin{equation}
g^{\mu \nu} \partial_{\mu} b \, \partial_{\nu} b=g^{tt} (\dot{b})^2+g^{rr} {b^{\prime}}^2=-\frac{\dot{b}^2}{a^2}+\frac{{b^{\prime}}^2}{b^2}.
\end{equation}
Thus, near the  $\Sigma_{1}$  surface, it is clear that $g^{\mu \nu} \partial_{\mu}b \,\partial_{\nu} b \ge 0$. Hence, $\Sigma_{1}$ surfaces are  timelike or null. The case $\delta^2-4 \gamma = 0$, representing only one singular dynamical surface, corresponds to the uncharged solutions which will be discussed in Section VI. For physical spacetimes both $\delta$ and $\gamma$ are positive. Hence in such cases $\Sigma_{1}$ surface does not exist.

\item[(ii)] \textbf{Surface $\Sigma_{2}$}

Regarding our definition for $q(t)$ as $a(t,r)=q(t)\frac{\dot{b}(t,r)}{b(t,r)}$ and since
\begin{equation}
\dot{b}(t,r)=\dot{\beta}(t) \left(\frac{1}{c_{0}+c_{1} r^2}-\frac{\gamma}{\beta^2}\, \frac{1}{c_{2}+c_{3} r^2} \right),
\end{equation}
then $\Sigma_{2}$ is defined as
\begin{equation}
X^2= \frac{c_{0}+c_{1} r^2}{c_{2}+c_{3} r^2}=\frac{\beta^2}{\gamma}.
\end{equation}
The normal vector of this surface satisfies
\begin{equation}
g^{\mu \nu} \partial_{\mu} a \, \partial_{\nu} a=g^{tt} \dot{a}^2+g^{rr} {a^{\prime}}^2=-\frac{\dot{a}^2}{a^2}+\frac{{a^{\prime}}^2}{b^2},
\end{equation}
representing that $\Sigma_{2}$ is a spacelike surface or null, since $a=0$ then $g^{\mu \nu} \partial_{\mu} a \partial_{\nu} a \leq0$ near $\Sigma_2$. Such singularities are named as the cosmological singularities \cite{penrose} or sudden cosmological singularities \cite{barrow1}-\cite{lake2004}.
\end{description}

\vspace{0.5cm}
\noindent
{\bf 2. Singular Surfaces for  the Class of $N=1$}\\
Regarding (\ref{n1bb}), the surface $\Sigma_{1}$ is given by
\begin{equation}
\nu_{0}(r) \left(b_{0}+b_{1} r^2\right)+\beta(t)=0,
\end{equation}
which is a timelike or null surface. In this case, there exists no $\Sigma_{2}$ surface.

\subsection*{B.~Properties of the Solution for $N=2$}
Our  new solution (\ref{qltc}) can be written as
\begin{equation} \label{bb22}
b(t,r)= \frac{\beta(t)}{c_{0}+c_{1}\,r^2}\,\left(1+\frac{\delta}{\beta(t)}\, \sqrt{\frac{c_{0}+c_{1}\,r^2}{c_{2}+c_{3}\,r^2}}+\frac{\gamma}{\beta^2(t)}\, \frac{c_{0}+c_{1}\,r^2}{c_{2}+c_{3}\,r^2} \right).
\end{equation}
Using (\ref{denk8}), (\ref{gfg}), (\ref{ggg}) and (\ref{eqn5}), the functions $h$ and $h_{1}$ can be obtained as
\begin{eqnarray}
&&h(r)= \frac{\sqrt{\delta^2-4\gamma}\,(c_{0}\, c_{3}-c_{1}\,c_{2})\,r}{(c_{0}+c_{1} r^2)^{3/2}\, (c_{2}+c_{3} r^2)^{3/2}}, \label{hh1}\\
&&h_{1}(r)=\, \frac{3 \delta(c_{0}\, c_{3}-c_{1}\,c_{2})^2 \,r^3}{(c_{0}+c_{1} r^2)^{5/2}\, (c_{2}+c_{3} r^2)^{5/2}}. \label{hh2}
\end{eqnarray}
Then, using (\ref{qt}) and (\ref{hh1}), the total charge $Q_T$ in a spherical region with the radius $R_0$ is
given by
\begin{equation}
Q_T=\frac{\sqrt{\delta^2 -4\gamma}\left( c_0c_3-c_1 c_2\right)R_0^3}{\left(c_0 +c_1 R_0^2  \right)^\frac{3}{2} \left(c_2 +c_3 R_0^2  \right)^\frac{3}{2}}.
\end{equation}

In our solution  $b(t,r)$ in (\ref{bb22}) there are six arbitrary constants. We can reduce this number to four by scaling.
It is easy to show that the function $b(t,r)$ is form invariant under the following scalings
\begin{eqnarray}
&&c_{0}=\frac{\bar{c}_{0}}{m},~c_{1}=\frac{\bar{c}_{1}}{m},~c_{2}=\frac{\bar{c}_{2}}{n},~c_{3}=\frac{\bar{c}_{3}}{n}, \\
&& \delta=\frac{\bar{\delta}}{\sqrt{m n}},~\gamma=\frac{\bar{\gamma}}{mn},~\beta=\frac{\bar{\beta}}{m}
\end{eqnarray}
where $m$ and $n$ are arbitrary nonzero real numbers. Hence, out of 6 parameters only 4 of them can be considered generic. In the next sections, without loosing any generality we use the following  two different parameterizations
to represent our new solution.
\begin{description}
\item[(A)] ~$c_{0}=1,~~c_{1}=k,~~ \frac{c_2}{c_{3}}=\mu$  where $\mu$ is any  real number.
\item[(B)] ~$c_{0}=1,~~ c_{1}=k_{1}, ~\frac{c_3}{c_{2}}=k_{2}$ where $k_{1}$ and $k_{2}$ are any real numbers.
\end{description}
Here in the case of part $A$ we will consider only the cases
$\mu >0,~ k_{1}>0,~ k_{2} >0$.
The reason for presenting the above two different representations of our solution is to show how it differs from the known exact solutions.

\vspace{0.5cm}
\noindent
{\bf A. The First Representation: The case of  $\mu=\frac{c_{2}}{c_{3}}$}\label{dafne000}\\
 For $c_{3} \ne 0$, we can consider the following identifications
\begin{eqnarray}
&&c_{0}=1,~~ c_{1}=k,~~\beta= a(t),\nonumber\\
&&\frac{\delta}{ \sqrt{c_{3}}}=M,~~\frac{4\gamma}{c_{3}}=M^2-Q^2,
~~\mu=\frac{c_{2}}{c_{3}},
\end{eqnarray}
where $c_3>0$ and $k$ is the spatial curvature constant corresponding to $0$
for the flat and to $\pm1$ for closed and open universes in general.  Using the above
 identifications and the $\delta$ in (\ref{ggg}), we can obtain our $c_4$ constant as
 \begin{equation}
c_4=\frac{{c_3} M^4}{2(1-k\mu)}, ~~~\mu k \ne 1 .
\end{equation}
We defined our constants $c_0, c_1, c_2, c_3$ and $c_4$ in such a way that
our solution (\ref{bb22}) reduces to Vaidya-Shah solution (\ref{bb122}) (for
either $c_0=0$
or $c_2=0$), as we will
see in Section V A 1.
Then, our  $a(t,r)$, $b(t,r)$, $h(r)$,  $h_1(r)$, $\sigma(t,r)$, $\rho(t,r)$, $p(t,r)$ and $F_{01}(t,r)$ functions become
\begin{eqnarray}
&&a(t,r)=\frac{1-\frac{M^2-Q^2}{4a^2(t)}\, \frac{1+kr^2}{\mu+r^2} }{\left(1+\frac{M}{a(t)}\, \sqrt{\frac{1+k\,r^2}{\mu+r^2}}+\frac{M^2-Q^2}{4a^2(t)}\, \frac{1+kr^2}{\mu+r^2} \right)} \label{1newa},\\
&&b(t,r)= \frac{a(t)}{1+k r^2}\,\left(1+\frac{M}{a(t)}\, \sqrt{\frac{1+k\,r^2}{\mu+r^2}}+\frac{M^2-Q^2}{4a^2(t)}\, \frac{1+kr^2}{\mu+r^2} \right), \label{1newb}\\
&&h(r)=\frac{2 |Q|\left(1-\mu k\right)r}{\left(1+k r^2\right)^{3/2}\, (\mu+r^2)^{3/2}}, \\
&&h_1(r)=\frac{6M\left(1-\mu k\right)^2 r^3}{\left(1+k r^2\right)^{5/2}\, (\mu+r^2)^{5/2}}, \\
&& F_{01}(t,r)=\psi(t,r)=h(r) \frac{a(t,r)}{b(t,r)},\\
&& 4 \pi \sigma(t,r)= \frac{3|Q| \left(1-\mu k\right)(\mu - k r^4)(1+k r^2)^{\frac{1}{2}}}{a^3(t)(\mu+r^2)^{\frac{5}{2}}\left(1+\frac{M}{a(t)}\, \sqrt{\frac{1+k\,r^2}{\mu+r^2}}+\frac{M^2-Q^2}{4a^2(t)}\, \frac{1+kr^2}{\mu+r^2} \right)^3 },\label{sigi}\\
&&8\pi \rho(t,r)=3\frac{\dot a^2(t)}{a^2(t)}-\frac{S(t,r)}{b^4(t,r)} -\Lambda,\label{rogi}\\
&&8\pi p(t,r)=-3\frac{\dot a^2(t)}{a^2(t)}+2\left(\frac{\dot a^2(t)}{a^2(t)}-\frac{\ddot a(t)}{a(t)}\right)X(t,r)+\frac{Y(t,r)}{\left( 1-\frac{M^2-Q^2}{4a^2(t)}\frac{1 +kr^2}{\mu +r^2} \right)b^4(t,r)}+\Lambda\label{pigi},
\end{eqnarray}
where $S(t,r),~ X(t,r)$ and $Y(t,r)$ functions are given in  Appendix B.
Here, without losing any generality, we have set $q(t) \dot a(t)/a(t)=1$. Hence, in our new solution (\ref{1newa})  and (\ref{1newb}), in addition to the mass $M$, charge $Q$ and the spatial curvature constant $k$, we have a new parameter $\mu$.  When $\mu=0$, this solution reduces to the Vaidya-Shah solution (\ref{bb122}), as we will
see in Section V A 1\ref{vs-section}. Our solution reduces to the Reissner-Nordstr{\" o}m metric when $\mu=k=0$ and $a(t)=1$  in isotropic coordinates. When $\mu=k=0$ and $a(t)=e^{\sqrt{\frac{\Lambda}{3}}\,t}$ then we obtain Schwarzschild-Reissner-Nordstr{\" o}m-de Sitter metric with cosmological constant $\Lambda$.

\vspace{0.5cm}
\noindent
{\bf Remark 1}:
We point out that in contrast to the Vaidya-Shah solution, in our new solution, the current vector $J^{\mu}$  (or the charge density $\sigma$) is non-zero for the flat spatial curvature constant, i.e. $k=0$. On the other hand if $\mu k=1$ where the charge density and the total charge in a volume of radius $R_{0}$ vanish our solution reduces to the FRW metric (see Remark 3).

\vspace{0.5cm}
\noindent
For this solution, we have the following points.

\begin{description}
\item[(i)] The surface $\Sigma_1$ is given as
\begin{equation}
\Sigma_1: \sqrt{\frac{\mu+r^2}{1+k r^2}}=\frac{-M \pm |Q|}{2 a(t)}.
\end{equation}
Hence, $\Sigma_{1}$ exists only when $M <0$ and $|Q| > M$ as we mentioned also in the general case in page 8.
Then, we conclude that $\Sigma_{1}$ does not exist for physical cases.
\item[(ii)] The surface $\Sigma_2$ is given by the following equation
\begin{equation}\label{siig}
\Sigma_2:~~(M^2-Q^2)(1+kr^2)-4a^{2}(t)(\mu+r^2)=0,
\end{equation}
which requires $M^2-Q^2>0$.
\item[(iii)] For the extreme case, i.e $M=|Q|$,  $\Sigma_{1}$ does not exist
and $\Sigma_2$ corresponds to $a(t)=0$ (The big-bang singularity).

\item[(iv)] At the spatial origin, i.e $r\rightarrow0$, the metric functions $a(t,r)$ and $b(t,r)$ as well as $\sigma(t,r)$, $\rho(t,r)$ and $p(t,r)$  are nonsingular in general except for the cosmological models
with $a(t)\rightarrow0$, see  Appendix B for more details.
\item[(v)] At the spatial infinity, i.e $r\rightarrow \infty$,  the metric functions $a(t,r)$ and $b(t,r)$ as well as $\sigma(t,r)$, $\rho(t,r)$ and $p(t,r)$  remains regular and the behavior of this model at the asymptotic region is different
than the FRW solution,   see  Appendix B for more details.
\end{description}

\vspace{0.5cm}
\noindent
{\bf B. The Second Representation: The case of $k_2=\frac{c_{3}}{c_{2}}$}\label{dafne0001}\\
For $c_{2} \ne 0$, one may also consider the following identifications
\begin{eqnarray}
&&c_{0}=1,~~ c_{1}=k_{1},~~\beta= a(t),\nonumber\\
&&\frac{\delta}{ \sqrt{c_{2}}}=M,~~\frac{4\gamma}{c_{2}}=M^2-Q^2,
~~k_2=\frac{c_{3}}{c_{2}},
\end{eqnarray}
where $c_2>0$ and $k_1$ and $k_2$ are two generally   different spatial curvatures. Using the above identifications and the $\delta$ in (\ref{ggg}), we can obtain $c_4$ constant as\begin{equation}
c_4=\frac{{c_2} M^4}{2(k_{2}-k_1)},
\end{equation}
where $k_1 \neq k_2$. For this case, the  $a(t,r)$, $b(t,r)$, $h(r)$, $h_1(r)$, $\sigma(t,r)$, $\rho(t,r)$, $p(t,r)$ and $F_{01}(t,r)$ functions can be found as
\begin{eqnarray}
&&a(t,r)=\frac{1-\frac{M^2-Q^2}{4a^2(t)}\, \frac{1+k_{1}r^2}{1+k_2r^2} }{\left(1+\frac{M}{a(t)}\, \sqrt{\frac{1+k_{1}\,r^2}{1+k_2r^2}}+\frac{M^2-Q^2}{4a^2(t)}\, \frac{1+k_{1}r^2}{1+k_{2}r^2} \right)} \label{newa},\\
&&b(t,r)= \frac{a(t)}{1+k_{1} r^2}\,\left(1+\frac{M}{a(t)}\, \sqrt{\frac{1+k_{1}\,r^2}{1+k_2
r^2}}+\frac{M^2-Q^2}{4a^2(t)}\, \frac{1+k_1 r^2}{1+k_2 r^2} \right), \label{newbp}\\
&&h(r)=\frac{2 |Q| (k_{2}- k_1)r}{(1+k_{1} r^2)^{\frac{3}{2}}\, (1+k_{2}r^2)^{\frac{3}{2}}}, \\
&&h_1(r)=\frac{6M\left(k_{2}-k_1\right)^2 r^3}{\left(1+k_{1} r^2\right)^{\frac{5}{2}}\, (1+k_2r^2)^{\frac{5}{2}}}, \\
&& F_{01}(t,r)=\psi(t,r)=h(r) \frac{a(t,r)}{b(t,r)},\\
&&4 \pi \sigma(t,r)= \frac{3 |Q| \left(k_2-k_1\right)(1 - k_1 k_2 r^4)(1+k_1 r^2)^{\frac{1}{2}}}{a^3(t)(1+k_2r^2)^{\frac{5}{2}}\left(1+\frac{M}{a(t)}\, \sqrt{\frac{1+k_1\,r^2}{1+k_2r^2}}+\frac{M^2-Q^2}{4a^2(t)}\, \frac{1+k_1r^2}{1+k_2 r^2} \right)^3 },\label{sen0}\\
&&8\pi \rho(t,r)=3\frac{\dot a^2 (t)}{a^2(t)}-\frac{S(t,r)}{b^4(t,r)} -\Lambda,\label{siz0}\\
&&8\pi p(t,r)=-3\frac{\dot a^2(t)}{a^2(t)}+2\left(\frac{\dot a^2(t)}{a^2(t)}-\frac{\ddot a(t)}{a(t)}\right)X(t,r)+\frac{Y(t,r)}{\left( 1-\frac{M^2-Q^2}{4a^2(t)}\frac{1 +k_1r^2}{1 +k_2r^2} \right)b^4(t,r)}+\Lambda,\label{olmaz0}
\end{eqnarray}
where $S(t,r),~ X(t,r)$ and $Y(t,r)$ functions are given in  Appendix C.

\vspace{0.5cm}
\noindent
{\bf Remark 2}: We point out that the case $k_{1}=k_{2}=k$ reduces to a FRW metric with spatial curvature $k$ (see Remark 3).

\vspace{0.5cm}
\noindent
For this solution, one realizes the following points.
\begin{description}
\item[(i)] Depending on the sign and values of $M\neq0$ and $Q\neq0$ parameters, we have
\begin{equation}
\Sigma_1: \sqrt{\frac{1+k_{2} r^2}{1+k_{1} r^2}}=\frac{-M \pm |Q|}{2 a(t)}.
\end{equation}
Hence, we have exactly the same conclusion as the previous case that $\Sigma_{1}$ does not exist for physical cases.
\item[(ii)] The surface $\Sigma_2$ is given by the following equation
\begin{equation}
\Sigma_2:~~(M^2-Q^2)(1+k_1r^2)-4a^{2}(t)(1+k_2r^2)=0,
\end{equation}
which requires $M^2-Q^2>0$.
\item[(iii)] For the extreme case, i.e $M=|Q|$, $\Sigma_{1}$ does not exist
and $\Sigma_2$ corresponds to $a(t)=0$ (The big-bang singularity).
\item[(iv)] At the spatial origin, i.e $r\rightarrow0$, the metric functions $a(t,r)$ and $b(t,r)$ as well as $\sigma(t,r)$, $\rho(t,r)$ and $p(t,r)$  are nonsingular in general except for the cosmological models
with $a(t)\rightarrow0$, see  Appendix C for more details.
\item[(v)] At the spatial infinity, i.e $r\rightarrow \infty$,  the metric functions $a(t,r)$ and $b(t,r)$ as well as $\sigma(t,r)$, $\rho(t,r)$ and $p(t,r)$  remains regular and the behavior of this model at the asymptotic region is different
than the FRW solution,   see  Appendix C for more details.
\end{description}

\vspace{0.3cm}
We summarize this section with the following Theorem.

\vspace{0.1cm}
\noindent

{\bf Theorem 3}: {\it The spacetime represented by our solution for $N=2$ either in (\ref{1newa}) and (\ref{1newb})  or in (\ref{newa}) and (\ref{newbp}) are nonsingular in  the sense that all the functions $a(t,r), b(t,r), p(t,r),
\rho(t,r)$ and $\sigma(t,r)$ either go to zero or to a finite value as $r$ goes to zero or to infinity.}
%

%
\subsection*{C.~Properties of the Solution for $N=1$}\label{n1case}
For $N=1$,  using (\ref{denk14}) and (\ref{denk15}), the functions $h$ and $h_1$ can
be obtained as
\begin{equation}
h(r)=\pm \left(\nu^{\prime}_{0}(r)-\nu_0(r)\frac{\nu^{\prime}_{1}(r)}{\nu_1(r)} \right),
\end{equation}
which can be written also as
\begin{eqnarray}
&&h(r)=\pm \left(\nu^{\prime}_{0}(r)+\nu_{0}(r) \frac{2 b_{1} r}{b_{0}+b_{1} r^2} \right).
\end{eqnarray}
Here, similar to the previous solutions and without losing any generality,
we set $\beta(t)=a(t)$, $~q(t)\dot a(t)/a(t)=1$, $b_{0}=1$ and  $b_{1}=k$.
Then, we have
\begin{eqnarray}
&&a(t,r)=\frac{1}{1+\frac{\nu_0(r)}{a(t)}(1+kr^2)}, \\
&& b(t,r)=\nu_{0}(r)+\frac{a(t)}{ 1+k\, r^2},\label{nsmkk}\\
&&h(r)=\pm \left(\nu^{\prime}_{0}(r)+\nu_{0}(r) \frac{2 k r}{1+k r^2} \right),\label{lrvs}\\
&&h_1(r)=r \nu_{0}^{\prime \prime}(r)- \nu_{0}^{\prime}(r)+2 r \frac{h^2(r)
-\nu^{\prime 2}_0(r)}{\nu_0(r)},\\
&&F_{01}(t,r)=\psi(t,r)=h(r) \frac{a(t,r)}{b(t,r)},\label{mast}\\
&&4 \pi \sigma(t,r)=\frac{1}{rb^3(t,r)}\left(rh^{\prime}(r)+2h(r)\right),\\
&&8\pi\rho(t,r)=3\frac{\dot a^2(t)}{a^2(t)}+\frac{1}{r b^4 } \left(-2 r b  b^{\prime\prime}- 4 b b^{\prime}+ r{b^{\prime}}^2-rh^2\right)-\Lambda, \\
&&8\pi p(t,r)=-3\frac{\dot a^2(t)}{a^2(t)}+\frac{2\dot q b}{q^3 \dot b}+\frac{1}{r  b^4  \dot b} \left(2b(rb^{\prime} +b) {\dot b}^{\prime}-r{\dot b} {b^{\prime}}^2 +rh^2\dot
b \right)+\Lambda. \label{presure2}
\end{eqnarray}
Using (\ref{qt}) the total charge $Q_{T}$ in a bounded region with $r=R_{0}$ is given by
\begin{equation}
Q_{T}=R_{0}^2 \,\,h(R_{0})=\pm R_{0}^2 \, \left(\nu_{0}^{\prime}(R_{0})+\nu_{0}(R_{0})\, \frac{2 k R_{0}}{1+k R_{0}^2}\right)
\end{equation}
For this solution, the behavior of the metric functions, charge
and energy densities as well as the pressure at the spatial origin or  asymptotic region
in general depend on the explicit form of the arbitrary function $\nu_0(r)$.
Then, without its explicit form, we can not discuss accurately on the properties of
the singular surfaces as well as the properties at the spatial origin or infinity.
 We will introduce some special subclasses
of this general solution in the sections V B and VI B, and discuss briefly how these
subclasses can be a reasonable physical solutions or not. For example, regarding
(\ref{nsmkk}), one can show that the second term in the pressure (\ref{presure2})
diverges at $r\rightarrow \infty$ for $\nu_0(r) \propto \frac{1}{r^n}$ with $n<2$.
One may argue that these types of solutions can not be   reasonable physical solutions regarding their divergence at the asymptotic region. As our next work, we will classify various possible subclasses for $N=1$ regarding the possible physical choices for the arbitrary  $\nu_0(r)$ function.

\vspace{1cm}
\section*{V.~Special Subclasses of the General Solutions and Their Properties}
In this section, we investigate some particular subclasses of our general
solutions as well as their properties.
\subsection*{A.~Subclasses of $N=2$}
\subsubsection*{1.~The case of either $c_{0}=0 $ but $c_{2} \ne 0$, or $c_{2}=0$ but $c_{0} \ne 0$} \label{vs-section}
Both these cases correspond to the Vaidya-Shah solution (\ref{bb122}). To show that, for example, we consider the case of $c_{2}=0$ but $c_{0} \ne 0$ which leads to
\begin{equation}\label{bab3}
b(t,r)= \frac{\beta(t)}{c_{0}+c_{1}\,r^2}\,\left(1+\frac{\delta}{\beta(t)\sqrt{c_3}r}\, \sqrt{c_{0}+c_{1}\,r^2}+\frac{\gamma}{\beta^2(t)}\, \frac{c_{0}+c_{1}\,r^2}{c_{3}\,r^2} \right),
\end{equation}
where with the identifications of $c_{0}=1,~c_{1}=k,~\beta=a(t),~\frac{\delta}{ \sqrt{c_{3}}}=M,~c_3>0$ and $\frac{4\gamma}{c_{3}}=M^2-Q^2$ takes the form of
\begin{equation}
b(t,r)=\frac{a(t)}{1+k r^2}\, \left[ 1+M \frac{\sqrt{1+k r^2}}{a(t)r}+(M^2-Q^2) \frac{1+k r^2}{4a^2(t)\,r^2} \right].
\end{equation}
Consequently, we can also find the metric function $a(t,r)$ as
\begin{equation}
a(t,r)=\frac{1- \frac{M^2-Q^2}{4a^2(t)\,r^2} \left(1+k r^2\right)}{1+\frac{M}{a(t)r}\sqrt{1+k r^2}+ \frac{M^2-Q^2}{4a^2(t)\,r^2}(1+k r^2)},
\end{equation}
where without losing any generality, we have set $q(t) \dot a(t)/a(t)=1$. We see that our metric functions  $a(t,r)$ and  $b(t,r)$ exactly reduce to the  Vaidya-Shah solution (\ref{bb122}). Then, the Vaidya-Shah solution can
be considered  as one of the  particular subclasses of our generalized solution (\ref{bb3}). Also, the $k$ parameter here is the spatial curvature constant which in general corresponds to zero
for the flat, and to $\pm1$ for the closed and open universes, respectively.

One can realize the following points about this solution.
\begin{description}
\item[(i)] Depending on the sign and values of $M\neq0$ and $Q\neq0$ parameters, we have
\begin{equation}
\Sigma_1:~~(M\mp|Q|)\sqrt{1+kr^2}+2a(t)r=0.
\end{equation}
which exists only for unphysical cases, i.e $M-|Q|<0$.

\item[(ii)] The surface $\Sigma_2$ is given by the following equation
\begin{equation}
\Sigma_2:~~(M^2-Q^2)(1+kr^2)-4a^{2}(t)r^2=0,
\end{equation}
which requires $M^2-Q^2>0$.
\item[(iii)] For the extreme case, i.e $M=|Q|$, $\Sigma_{1}$ does not exist
and $\Sigma_2$ corresponds to $a(t)=0$ (The big-bang singularity).

\item[(iv)] As it is proved for the general solutions in the section \ref{dafne000}, for $r\rightarrow 0$ and $r\rightarrow \infty$, charge
density, mass density and pressure remain finite also for this subclass. \end{description}
\subsubsection*{2.~The case of either $c_{1}=0$ or $c_{3}=0$.} \label{chingili}
These cases correspond to the same spacetime geometry.
Then, we discuss only the case of  $c_{1}=0$  as follows.
For this case, our solution (\ref{bb3}), takes the following form
\begin{equation}\label{baaab3}
b(t,r)= \frac{\beta(t)}{c_{0}}\,\left(1+\frac{\delta}{\beta(t)}\, \sqrt{\frac{c_0}{c_{2}+c_{3}\,r^2}}+\frac{\gamma}{\beta^2(t)}\, \frac{c_{0}}{c_2 +c_{3}\,r^2} \right).
\end{equation}
Using (\ref{denk8}), (\ref{gfg}), (\ref{ggg}) and (\ref{eqn5}), the functions $h(r)$ and $h_{1}(r)$ can be obtained as
\begin{eqnarray}\label{hh1an}
&&h(r)=\sqrt{\frac{\delta^2-4\gamma}{c_0}} \frac{c_{3}r}{(c_{2}+c_{3} r^2)^{\frac{3}{2}}}, \\
&&h_{1}(r)=\, \frac{3 \delta c_{3}^2 \,r^3}{\sqrt{c_{0}} (c_{2}+c_{3} r^2)^{\frac{5}{2}}}.
\end{eqnarray}
Similar to the case of Vaidya-Shah solution, we can consider the following identifications
\begin{eqnarray}
&&\beta(t)= a(t), ~\,c_0=1,\nonumber\\
&&\frac{\delta}{\sqrt{c_{2}}}=M,~~\frac{4\gamma }{c_{2}}=M^2-Q^2,
~~k=\frac{c_{3}}{c_{2}},
\end{eqnarray}
where requires $c_2>0$. Then, the metric function  $b(t,r)$ takes the following form
\begin{equation}\label{baaab3v}
b(t,r)= a(t)\,\left(1+\frac{M}{a(t)}\, \frac{1}{\sqrt{1+kr^2}}+\frac{M^2-Q^2}{4a^2(t)}\, \frac{1}{1+kr^2} \right),
\end{equation}
where $k$ is the spatial curvature constant. Using the above
 identifications and the $\delta$ in (\ref{ggg}), we obtain our $c_4$ constant as
 
 \begin{equation}
c_4=\frac{{c_2} M^4}{2k},~~ k \ne 0.
\end{equation}
Then, our  $a(t,r)$, $b(t,r)$, $h(r)$,  $h_1(r)$, $\sigma(t,r)$, $\rho(t,r)$, $p(t,r)$ and $F_{01}(t,r)$ functions become
\begin{eqnarray}
&&a(t,r)=\frac{1-\frac{M^2-Q^2}{4a^2(t)}\, \frac{1}{1+kr^2} }{\left(1+\frac{M}{a(t)}\, \frac{1}{\sqrt{1+kr^2}}+\frac{M^2-Q^2}{4a^2(t)}\, \frac{1}{1+kr^2} \right)} \label{nea},\\
&&b(t,r)= a(t)\,\left(1+\frac{M}{a(t)}\, \frac{1}{\sqrt{1+kr^2}}+\frac{M^2-Q^2}{4a^2(t)}\, \frac{1}{1+kr^2} \right), \label{neb}\\
&&h(r)=\frac{ |Q| k\,r}{ (1+kr^2)^{\frac{3}{2}}}, \\
&&h_1(r)=\frac{3M\,k^2\, r^3}{ (1+kr^2)^{\frac{5}{2}}}, \\
&& F_{01}(t,r)=\psi (t,r)=h(r) \frac{a(t,r)}{b(t,r)}\\
&& 4 \pi \sigma(t,r)=\frac{3 k|Q|}{ a^{3}(t)\,(1+kr^2)^{\frac{5}{2}}\left(1+\frac{M}{a(t)}\, \frac{1}{\sqrt{1+kr^2}}+\frac{M^2-Q^2}{4a^2(t)}\, \frac{1}{1+kr^2} \right)^3
},\label{okan}\\
&&8\pi\rho(t,r)=3\frac{\dot a^2(t)}{a^2(t)}+\frac{S(t,r)}{b^4(t,r)}-\Lambda,\label{tek}\\
&&8\pi p(t,r)=-3\frac{\dot a^2(t)}{a^2(t)}+2\left(\frac{\dot a^2(t)}{a^2(t)}-\frac{\ddot a(t)}{a(t)}\right)X(t,r)+\frac{Y(t,r)}{\left( 1-\frac{M^2-Q^2}{4a^2(t)}\, \frac{1}{1+kr^2} \right)b^4(t,r)}+\Lambda,\label{men}
\end{eqnarray}
where $S(t,r),~ X(t,r)$ and $Y(t,r)$ functions are given in Appendix D. Here, without losing any generality, we have set $q(t) \dot a(t)/a(t)=1$. One can realize the following points about this solution.
\begin{description}
\item[(i)] Depending on the sign and values of $M\neq0$ and $Q\neq0$ parameters, we have
\begin{equation}
\Sigma_1:~~M\mp|Q|+2a(t)\sqrt{1+kr^2}=0,
\end{equation}
which exists only for unphysical cases, i.e $M-|Q|<0$.

\item[(ii)]  The surface $\Sigma_2$ is given by the following equation
\begin{equation}
\Sigma_2:~~(M^2-Q^2)-4a^{2}(t)(1+kr^2)=0,
\end{equation}
which requires $M^2-Q^2>0$.
\item[(iii)]  For the extreme case, i.e $M=|Q|$, $\Sigma_{1}$ does not exist
and $\Sigma_2$ corresponds to $a(t)=0$ (big-bang singularity).

\item[(iv)]  Again, as it is proved for the general solutions in the section IV B, for $r\rightarrow 0$ and $r\rightarrow \infty$, charge
density, mass density and pressure remain finite for this subclass. \end{description}

\subsubsection*{3.~The case of $c_{4}=0$ }
Regarding (\ref{ggg}), this case corresponds to $\delta=0$ and $\alpha_0=0$.
 Then, the metric function $b(t,r)$ in (\ref{bb3}) takes the following form
 \begin{equation}\label{bb344}
b(t,r)=\frac{\beta(t)}{ c_{0}+c_{1}r^2}\left(1+\frac{\gamma}{\beta^2(t)}\,\frac{c_{0}+c_{1}r^2}{c_{2}+c_{3}\, r^2}\right),
\end{equation}
 which similar to the sections IV A and IV B, can be demonstrated in both the $(k, \mu)$ and $(k_1, k_2)$ representations.
Also, we find that $h_1=0$, and the function $h(r)$ takes the following form
\begin{equation}
h(r)= \frac{\sqrt{-4\gamma}\,(c_{0}\, c_{3}-c_{1}\,c_{2})\,r}{(c_{0}+c_{1} r^2)^{3/2}\, (c_{2}+c_{3} r^2)^{3/2}},
\end{equation}
where requires the condition $\gamma<0$. One may consider $4\gamma=-c_2Q^2$
which reduces our solution here to the solutions with $M=0$ in  the sections
IV A and IV B, i.e to the charged massless solutions.
We consider   the following identifications
\begin{eqnarray}
&&c_{0}=1,~~ c_{1}=k_{1},~~\beta= a(t),\nonumber\\
&&~~\frac{4\gamma}{c_{2}}=-Q^2,
~~k_2=\frac{c_{3}}{c_{2}},
\end{eqnarray}
where here $k_1$ and $k_2$ are generally  two different spatial curvatures.  For this case,  $a(t,r)$, $b(t,r)$, $\sigma(t,r)$, $\rho(t,r)$, $p(t,r)$ and $F_{01}(t,r)$ functions read as\begin{eqnarray}
&&a(t,r)=\frac{1+\frac{Q^2}{4a^2(t)}\, \frac{1+k_{1}r^2}{1+k_2r^2} }{\left(1-\frac{Q^2}{4a^2(t)}\, \frac{1+k_{1}r^2}{1+k_{2}r^2} \right)} \label{newab},\\
&&b(t,r)= \frac{a(t)}{1+k_{1} r^2}\,\left(1-\frac{Q^2}{4a^2(t)}\, \frac{1+k_1 r^2}{1+k_2 r^2} \right), \label{newbj}\\
&&h(r)=\frac{ |Q|(k_{2}- k_1)r}{(1+k_{1} r^2)^{\frac{3}{2}}\, (1+k_{2}r^2)^{\frac{3}{2}}}, \\
&& F_{01}(t,r)=\psi(t,r)=h(r) \frac{a(t,r)}{b(t,r)},\\
&&4 \pi \sigma(t,r)= \frac{3 |Q|  \left(k_2-k_1\right)(1 - k_1 k_2 r^4)(1+k_1 r^2)^{\frac{1}{2}}}{a^3(t)(1+k_2r^2)^{\frac{5}{2}}\left(1-\frac{Q^2}{4a^2(t)}\, \frac{1+k_1r^2}{1+k_2 r^2} \right)^3 },\label{sen0n}\\
&&8\pi \rho(t,r)=3\frac{\dot a^2 (t)}{a^2(t)}-\frac{S(t,r)}{b^4(t,r)} -\Lambda,\label{siz0b}\\
&&8\pi p(t,r)=-3\frac{\dot a^2(t)}{a^2(t)}+2\left(\frac{\dot a^2(t)}{a^2(t)}-\frac{\ddot a(t)}{a(t)}\right)X(t,r)+\frac{Y(t,r)}{\left( 1+\frac{Q^2}{4a^2(t)}\frac{1 +k_1r^2}{1 +k_2r^2} \right)b^4(t,r)}+\Lambda,\label{olmaz0b}
\end{eqnarray}
where we have supposed $c_2>0$ and $S(t,r),~ X(t,r)$ and $Y(t,r)$ functions are given in Appendix E. Regarding (\ref{newab}) and (\ref{newbj}),
this solution is the generalization of Vaidya-Shah solution to the case of
two spatial curvature with $M=0$.

For this solution, one realizes the following points.
\begin{description}
\item[(i)]  The surface $\Sigma_1$ is given by the following equation
\begin{equation}
\Sigma_1: ~~~~-Q^{2}(1+k_{1}r^2)+4a^{2}(t)(1+k_2r^{2})=0,
\end{equation}
where in contrast to the previous cases, it does exist as a physical case.
\item[(ii)]  The surface $\Sigma_2$ is given by the following equation
\begin{equation}
\Sigma_2: ~~~~Q^2(1+k_1r^2)+4a^{2}(t)(1+k_2r^{2})=0,
\end{equation}
where can not exist as a physical case.
\item[(iii)]  Similarly, as it is proved for the general solutions in the section IV B, for $r\rightarrow 0$ and $r\rightarrow \infty$, charge
density, mass density and pressure remain finite for this subclass. \end{description}

\subsubsection*{4.~The case of $\gamma=0$}
Regarding the condition to obtain (\ref{bb3}), i.e. $\beta_{1}(t)\beta_{2}(t)=\gamma$,
 this case corresponds to the situation where at least one of $\beta_1(t)$ and $\beta_2(t)$ in (\ref{4d}) is zero. Then, this case reduces to the solution $N=1$ as in (\ref{bigul})
in the section III B.

\subsection*{B.~ Subclass of $N=1$}\label{n1sub}
\subsubsection*{1.~The case of $\nu_0=constant$}
For this case, our metric functions take the following form
\begin{eqnarray}
&&a(t,r)=\frac{1}{1+\frac{\nu_0}{a(t)}(1+kr^2) }, \\
&& b(t,r)=\nu_0+\frac{a(t)}{ 1+k\, r^2},
\end{eqnarray}
as well as
\begin{eqnarray}
h(r)&=&\pm  \frac{2\nu_0 k r}{1+k r^2}  ,\\
h_1 (r)&=&\frac{8\nu_0^2 k^2 r^3}{(1+k r^2)^2},\\
F_{01}(t,r)&=&\psi(t,r)=h(r) \frac{a(t,r)}{b(t,r)}, \\
4 \pi \sigma(t,r)&=&\pm  \frac{2 k\nu_0 (3+kr^{2})(1+kr^{2})}{(\nu_0(1+kr^2)
+a(t))^3},\label{seni}\\
8\pi\rho(t,r)&=&3\frac{{\dot a}^2(t)}{a^2(t)}+\frac{\left(1+kr^2\right)^4}{\left(\nu_0(1+kr^2)
+a(t)\right)^4 } \left(\frac{12\nu_0a^2(t)k}{\left(1+kr^2\right)^4} +\frac{4k\nu_0a(t)(3-kr^2)}{\left(1+kr^2\right)^3} -\frac{4\nu_0^2k^2r^2}{\left(1+kr^2\right)^2} \right)-\Lambda,\nonumber\\\label{meni3}
&&\\
8\pi p(t,r)&=&-3\frac{{\dot a}^2(t)}{a^2(t)}+2\left(\frac{\dot a^2(t)}{a^2(t)}-\frac{\ddot a(t)}{a(t)}\right)\left(1+\frac{\nu_0(1+kr^2)}{a(t)} \right)\nonumber\\
&&+\frac{4k\nu_{0}(1+kr^2)^2+8k\nu_0a(t)(1+kr^2)+4ka^2(t)}{\left( a(t)+\nu_0(1+kr^2)\right)^4}+\Lambda.\label{bizi3}
\end{eqnarray}
Then, one can find that for $r\rightarrow 0$, all the quantities in this solution are regular  while at the asymptotic region, i.e  $r\rightarrow \infty$, the pressure diverges
by its second term in (\ref{bizi3}). Then, regarding this unusual asymptotic behavior, one may argue that this solution can not be a physical charged
solution. However,  in the section VI B 2, we will show that the
solution for $\nu_0=constant$ can be a physical uncharged solution for the
flat universe ($k=0$). Also, as we stated at the end of the section IV C,  we will classify the
possible choices by this kind of physical arguments in our next work.

\section*{VI.~Uncharged Solutions and Their Properties}\label{unimani}
In this section, we explore the uncharged solutions and their properties
for $N=2$ and $N=1$ in detail.
\subsection*{A.~Uncharged Solutions for $N=2$}
To obtain the uncharged solutions  for $N=2$, regarding (\ref{psi2}) and
(\ref{hh1}), we first
assume that the constants $c_{0}$, $c_{1}$, $c_{2}$ and $c_{3}$ are nonzero. Then, we investigate special cases where some of these parameters vanish or they are related.  Regarding (\ref{psi2}) and (\ref{hh1}), there are two main possibilities
to obtain uncharged solutions.
\subsubsection*{1.~The case of $c_0 c_3=c_1 c_2$}\label{section611}

For this case,  the functions $h(r)$ and $F_{01}=0$ in (\ref{hh1}) and (\ref{psi2}),
 respectively, (as well as $h_1(r)$ in (\ref{hh2}) and $\delta$ in (\ref{ggg})) vanish. As a specific case, using  the identification of $\beta(t)=a(t)$,
$c_0=1$, $c_1=k$, $\frac{\gamma}{c_2}=M$ and then $\frac{\gamma k}{M}=c_3$ in (\ref{bb3}), we obtain
\begin{eqnarray}
&&a(t,r)=\frac{1-\frac{M}{a^2(t)}}{1+\frac{M}{a^2(t)} },\\
&&b(t,r)=\frac{a(t)}{ 1+k\,r^2}\left(1+\frac{M}{a^{2}(t)} \right).
\end{eqnarray}
Then, the spacetime metric becomes
\begin{equation}
ds^2=-a_{1}^2(t)\, dt^2+\frac{a_{2}^2(t)}{(1+k r^2)^2}\,\left(dr^2+r^2\, d\Omega^2\right), \label{frw1}
\end{equation}
where
\begin{equation}
a_{1}(t)=\frac{1-\frac{M}{a^2(t)}}{1+\frac{M}{a^2(t)} },~~~a_{2}(t)=a(t)\,\left(1+\frac{M}{a^{2}(t)} \right).
\end{equation}
By the coordinate transformations
\[
a_{1}(t) dt= dT,~~~\frac{r}{1+k r^2}=R,
\]
the new metric in the new coordinates $T$ and $R$ becomes
\begin{equation}
ds^2=-dT^2+\bar{a}_{2}^2(T)\, \left[\frac{dR^2}{1-4 k R^2}+R^2\, d\Omega^2 \right]. \label{frw2}
\end{equation}
where $\bar{a}_{2}(T)=a_{2}(t(T))$. Hence, this special case is identical
to the Friedmann-Robertson-Walker model.


\subsubsection*{2.~The case of $\delta^2=4 \gamma$}

For this case,  the function $h(r)$ in (\ref{hh1}) and consequently the
 function $F_{01}=\psi(t,r)$
 in (\ref{psi2}) vanish and the uncharged case ($\sigma(t,r)=0$) can be provided. Considering $\delta^2=4\gamma$, the metric function
  $b(t,r)$  in (\ref{bb3})   takes the form
of\begin{eqnarray}\label{bb23}
b(t,r)&=&\frac{\delta}{\sqrt{c_{0}+c_{1}\,r^2}\,\sqrt{c_{2}+c_{3}\,r^2}}+\frac{\beta(t)}{c_{0}+c_{1}\, r^2}+\frac{\delta^2}{4 \beta(t)}\,\frac{1}{c_{2}+c_{3}\, r^2} \nonumber \\
&=&\left(\frac{\delta}{2\, \sqrt{\beta(t)}}\, \frac{1}{\sqrt{c_{2}+c_{3}\,r^2}}+\frac{\sqrt{\beta(t)}}{\sqrt{c_{0}+c_{1}\,r^2}} \right)^2,
\end{eqnarray}
and the corresponding $a(t,r)$ metric function will be
\begin{equation}
a(t,r)=\frac{q(t) \dot{\beta}(t) \left(\frac{1}{c_{0}+c_{1}\, r^2}-\frac{\delta^2}{4 \beta^{2}(t)}\,\frac{1}{c_{2}+c_{3}\, r^2}\right)}{\left(\frac{\delta}{2\, \sqrt{\beta(t)}}\, \frac{1}{\sqrt{c_{2}+c_{3}\,r^2}}+\frac{\sqrt{\beta(t)}}{\sqrt{c_{0}+c_{1}\,r^2}} \right)^2}=
\frac{q(t) \dot{\beta}(t)}{\beta(t)}\, \frac{ \left(\frac{1}{\sqrt{c_{0}+c_{1}\, r^2}}-\frac{\delta}{2 \beta(t)}\,\frac{1}{\sqrt{c_{2}+c_{3}\, r^2}}\right)}{\left(\frac{1}{\sqrt{c_{0}+c_{1}\,r^2}} +\frac{\delta}{2\, \beta(t)}\, \frac{1}{\sqrt{c_{2}+c_{3}\,r^2}}\right)}.\label{aa1}
\end{equation}
Similar to the previous solutions,
one can  set  $\beta(t)=a(t)$ and $q(t)\dot a(t)/a(t)=1$. Here, we assume that  $a(t)$ is nonnegative for all $t$. Thus, we find
\begin{eqnarray}
&&b(t,r)=\left(\frac{\delta}{2\, \sqrt{a(t)}}\, \frac{1}{\sqrt{c_{2}+c_{3}\,r^2}}+\frac{\sqrt{a(t)}}{\sqrt{c_{0}+c_{1}\,r^2}} \right)^2   \label{gfb1},\\
&&a(t,r)=\frac{1-\frac{\delta}{2 a(t)}\,\sqrt{\frac{c_{0}+c_{1}\, r^2}{c_{2}+c_{3}\, r^2}}}{1+\frac{\delta}{2\, a(t)}\, \sqrt{\frac{c_{0}+c_{1}\, r^2}{c_{2}+c_{3}\, r^2}}} \label{gfb},\\
&&h_{1}(r)=\, \frac{3 \delta(c_{0}\, c_{3}-c_{1}\,c_{2})^2 \,r^3}{(c_{0}+c_{1} r^2)^{\frac{5}{2}}\, (c_{2}+c_{3} r^2)^{\frac{5}{2}}},\\
&&8\pi \rho(t,r)=3\frac{{\dot a}^2(t)}{a^2(t)}-\frac{1}{ rb^{4}}\left(3 r  {b^{\prime}}^2+6 bb^{\prime}+2 h_{1}b \right)-\Lambda,\label{dfzeraer}\\
&&8\pi p(t,r)=-3\frac{{\dot a}^2(t)}{a^2(t)}+2\frac{b}{\dot b}\left(\frac{{\dot
a}^3(t)}{a^3(t)}-\frac{\ddot
a (t) \dot a(t)}{a^2(t)} \right)\\
&& ~~~~~~~~~~~~~~~+\frac{1}{ r b{^4} \dot
b}\, \left(2 b \left(r b^{\prime}+b\right)\, {\dot b}^{\prime}
- r{\dot b}\, {b^{\prime}}^2\right)+\Lambda\label{pba}.
\end{eqnarray}
 We have the following subclasses of  this general solution.
Here also we define our $c_0, c_1, c_2$ and  $c_3$ parameters in such a way that our general solution reduces
to the McVittie solution as one of its particular subclasses.
\\
\begin{description}
\item[(1)] \textbf{The case of $c_{1}=c_{2}=0$ or $c_{0}=c_{3}=0$}.

For this case, the metric functions (\ref{gfb1}) and (\ref{gfb}) take the following forms
\begin{eqnarray}
&&a(t,r)=\frac{1-\frac{\delta}{2 a(t)r}\,\sqrt{\frac{c_{0}}{c_{3}}}}{1+\frac{\delta}{2\, a(t)r}\, \sqrt{\frac{c_{0}}{c_{3}}}},\\
&&b(t,r)=\left(\frac{\delta}{2\, \sqrt{a(t)}}\, \frac{1}{\sqrt{c_{3}}r}+\frac{\sqrt{a(t)}}{\sqrt{c_{0}}} \right)^2,
\end{eqnarray}
where by the identifications $c_{0}=1$, and $\sqrt{c_{3}}=\frac{\delta}{M}$,
they read as
\begin{eqnarray}
&&a(t,r)=\frac{1-\frac{M}{2a(t)r}}{1+\frac{M}{2a(t)r}} ,\\
&&b(t,r)= a(t)\left(1+\frac{M}{2\, a(t) r}  \right)^2.
\end{eqnarray}
This solution is  the McVittie solution  for the flat
background universe $(k=0)$. The density and pressure profiles of this case
can be obtained as
\begin{eqnarray}
&&8\pi \rho(t,r)=3\frac{\dot a^2(t)}{a^2(t)}-\Lambda,\\
&&8\pi p(t,r)=-3\frac{\dot a^2(t)}{a^2(t)}+2\left(\frac{\dot a^2(t)}{a^2(t)}-\frac{\ddot a(t)}{a(t)}  \right)\frac{1+\frac{M}{2a(t)r}}{1-\frac{M}{2a(t)r}}+\Lambda.
\end{eqnarray}
Then, one finds the following points for this solution.
\begin{description}
\item[(a)] $\Sigma_1$ surface exists only for $M<0$ as
\begin{equation}
\Sigma_1:~~M+2a(t)r=0.
\end{equation}
\item[(b)]  $\Sigma_2$ surface
exists only for $M>0$ given by
\begin{equation}
\Sigma_2:~~M-2a(t)r=0.
\end{equation}
Interestingly, one notes that regarding the sign of $M$ parameter, the singular
surfaces can be spacelike or timelike.
However, for physical cases $M>0$, the only existing singular surface is
the spacelike surface $\Sigma_2$. This singular surface corresponds to the
big-bang singularity  \cite{kaloper}.
\end{description}
\item[(2)] \textbf{The case of $c_{2}=0$ or $c_0=0$}.\\
For the case $c_{2}=0$, the metric functions (\ref{gfb1}) and (\ref{gfb}) take the following forms
\begin{eqnarray}
&&a(t,r)=\frac{1-\frac{\delta}{2 a(t)r}\,\sqrt{\frac{c_{0}+c_1 r^2}{c_{3}}}}{1+\frac{\delta}{2\, a(t)r}\, \sqrt{\frac{c_{0}+c_1 r^2}{c_{3}}}},\\
&&b(t,r)=\left(\frac{\delta}{2\, \sqrt{a(t)}}\, \frac{1}{\sqrt{c_{3}}r}+\frac{\sqrt{a(t)}}{\sqrt{c_{0}+c_1
r^2}} \right)^2,
\end{eqnarray}
where by the identifications $c_{0}=1$, $c_{1}=k$ and $\frac{\delta}{\sqrt{c_{3}}}=M$,
 we have
\begin{eqnarray}
&&a(t,r)=\frac{1-\frac{M}{2a(t)r}\sqrt{1+kr^2}}{1+\frac{M}{2a(t)r}\sqrt{1+kr^2}} ,\\
&&b(t,r)= \frac{a(t)}{1+k r^2}\left(1+\frac{M}{2\, a(t) r} \sqrt{1+k r^2} \right)^2.
\end{eqnarray}
This solution is the generalization of the McVittie solution to a non-flat
background universe $(k\neq0)$. This solution can be also identified to the Vaidya-Shah
solution with $Q=0$, where we have previously addressed its asymptotic behavior
in section \ref{vs-section}.
Then, one finds the following points for this solution.
\begin{description}
\item[(a)] $\Sigma_1$ surface is given by \begin{equation}
\Sigma_1:~~M\sqrt{1+kr^2}+2a(t)r=0,
\end{equation}
which exists only for the unphysical cases, i.e for $M<0$.

\item[(b)]  $\Sigma_2$ surface
exists  for  $M>0$ as
\begin{equation}
\Sigma_2:~~M\sqrt{1+kr^2}-2a(t)r=0.
\end{equation}
\end{description}
\item[(3)] \textbf{The case of $c_{1}=0$ or $c_{3}=0$.}

For this case $c_{1}=0$, the metric functions (\ref{gfb1}) and (\ref{gfb}) take the following forms
\begin{eqnarray}
&&a(t,r)=\frac{1-\frac{\delta}{2 a(t)}\,\sqrt{\frac{c_{0}}{c_2+c_{3}r^2}}}{1+\frac{\delta}{2\, a(t)}\, \sqrt{\frac{c_{0}}{c_2+c_{3}r^2}}},\\
&&b(t,r)=\left(\frac{\delta}{2\, \sqrt{a(t)}}\, \frac{1}{\sqrt{c_2+c_{3}r^2}}+\frac{\sqrt{a(t)}}{\sqrt{c_{0}}} \right)^2,
\end{eqnarray}
where by the identifications $c_{0}=1$, $\frac{c_3}{c_{2}}=k$ and $\frac{\delta}{\sqrt{c_{2}}}=M$,
they read as \begin{eqnarray}\label{bbdfcx}
&&a(t,r)=\frac{1-\frac{M}{2 a(t)}\,\frac{1}{\sqrt{1+kr^2}} }{1+\frac{M}{2\, a(t)}\, \frac{1}{\sqrt{1+kr^2}}} ,\\
&&b(t,r)= a(t)\left(1+\frac{M}{2\, a(t)\sqrt{1+kr^2} }  \right)^2.
\end{eqnarray}
This solution is also another generalization of the McVittie solution to a non-flat background universe $(k\neq0)$ with different identification set
of our integration constant parameters.  We have studied the charged generalization
of this solution in section VA 2 with detail of its behavior at the spatial origin and infinity. Then, to avoid the repetition, one can set $Q=0$
to realize the properties of this solution.
Then, one finds the following points for this solution.
\begin{description}
\item[(a)] $\Sigma_1$ surface exists only for $M<0$ as
\begin{equation}
\Sigma_1:~~M+2a(t)\sqrt{1+kr^2}=0.
\end{equation}
\item[(b)]  $\Sigma_2$ surface
exists  for  $M>0$ as

\begin{equation}
\Sigma_2:~~~~M-2a(t)\sqrt{1+kr^2}=0.
\end{equation}
\end{description}
\item[(4)] \textbf{The case where none of the $c_i$ parameters are zero.}

For this case, the metric functions (\ref{gfb1}) and (\ref{gfb}) take the following forms
\begin{eqnarray}
&&a(t,r)=\frac{1-\frac{\delta}{2 a(t)}\,\sqrt{\frac{c_{0}+c_{1}\, r^2}{c_{2}+c_{3}\, r^2}}}{1+\frac{\delta}{2\, a(t)}\, \sqrt{\frac{c_{0}+c_{1}\, r^2}{c_{2}+c_{3}\, r^2}}},\\
&&b(t,r)=\left(\frac{\delta}{2\, \sqrt{a(t)}}\, \frac{1}{\sqrt{c_{2}+c_{3}\,r^2}}+\frac{\sqrt{a(t)}}{\sqrt{c_{0}+c_{1}\,r^2}} \right)^2,
\end{eqnarray}
 where by  identifications $c_{0}=1$, $c_{1}=k_1$, $\sqrt{c_{2}}=\frac{\delta}{M}$ and $k_2=\frac{c_{3}}{c_{2}}$, they read as
\begin{eqnarray}\label{bbvccx}
&&a(t,r)=\frac{1-\frac{M}{2 a(t)}\,\sqrt{ \frac{1+k_{1} r^2}{1+k_{2}r^2}} }{1+\frac{M}{2\, a(t)}\, \sqrt{ \frac{1+k_{1} r^2}{1+k_{2}r^2}}},\\
&&b(t,r)=\frac{M}{\sqrt{1+k_{1}\,r^2}\,\sqrt{1+k_{2}\,r^2}}+\frac{a(t)}{1+k_{1}\, r^2}+\frac{M^2}{4 a(t)}\,\frac{1}{1+k_{2}\, r^2}\\
&& ~~~~~~~~= \left(\frac{M}{2 \sqrt{a(t)}} \frac{1}{\sqrt{1+k_{2}\,r^2}}+ \frac{\sqrt{a(t)}}{\sqrt{1+k_{1} r^2}} \right)^2.\label{kcr}
\end{eqnarray}
{\bf Remark 3}:  When the metric function $b(t,r)$ takes the form
\[
b(t,r)=\frac{R(t)}{1+k r^2},
\]
we showed that, in Section VI A 1, that the spacetime reduces to the FRW universe with scale factor $R(t)$ and spatial curvature  $4 k$. Hence this suggests us that the metric function $b(t,r)$ in (\ref{kcr}) is a kind of nonlinear superposition of two different FRW $b$-functions
\[
b(t,r)=(b_{1}(t,r)+b_{2}(t,r))^2,
\]
where
\[
b_{1}(t,r)=\frac{\sqrt{a(t)}}{\sqrt{1+k_{1} r^2}},~~~b_{2}(t,r)=\frac{M}{2 \sqrt{a(t)}} \frac{1}{\sqrt{1+k_{2} r^2}}.
\]
If $k_{1} \ne k_{2}$ each one describes different FRW universes. The function $b_{1}(t,r)$ belongs to a FRW universe with the scale factor $a(t)$ and the spatial curvature $4k_{1}$ and the function $b_{2}(t,r)$ belongs to another FRW universe with the scale factor  $\frac{M^2} {4 a(t)}$ and the spatial curvature $4 k_{2}$. Hence our uncharged solution is a kind of a nonlinear superposition of two different FRW metrics with different spatial curvatures. If initially $a(t) \to 0$ then the function $b_{2}(t,r)$  is dominant in $b(t,r)$ and the corresponding universe is initially a FRW universe with spatial curvature $4 k_{2}$, see Appendix F. On the other hand if $a(t) \to \infty$ as $t \to \infty$ then the function $b_{1}(t,r)$ is dominant in the function $b(t,r)$  and the universe is described by a FRW metric with the spatial curvature $4 k_{1}$, see Appendix F. If $k_{1} k_{2} \le 0$ then we obtain an interesting result saying that the universe undergoes a kind of a topological change. In between, for $t \in (0, \infty)$, the  universe is a mixture of the above two FRW universes. If $k_{1}=k_{2}=k$ then the two FRW universes collapse to a single one with the spatial curvature $4k$.

One realizes the following points for this solution.
\begin{description}
\item[(i)] $\Sigma_1$ surface exists only for $M<0$ as
\begin{equation}
\Sigma_1:~~M\sqrt{1+k_1r^2}+2a(t)\sqrt{1+k_2r^2}=0.
\end{equation}
\item[(ii)]  $\Sigma_2$ surface
exists  for  $M>0$ as

\begin{equation}
\Sigma_2:~~M\sqrt{1+k_1r^2}-2a(t)\sqrt{1+k_2r^2}=0.
\end{equation}
\end{description}
\end{description}

\subsubsection*{3.~The case of $c_0=c_2=0$}
For this case, $\delta$, $h$ and $h_1$ functions vanish and the metric function $b(t,r)$ reads as
\begin{equation}
b(t,r)=\frac{\beta(t)}{c_1 r^2}\left(1+\frac{\gamma}{\beta^2(t)}\frac{c_1}{c_3} \right),
\end{equation}
where by defining $\tilde a(t)=\frac{\beta(t)}{c_1}\left(1+\frac{\gamma}{\beta^2(t)}\frac{c_1}{c_3} \right)$ takes the following simple form
\begin{equation}
b(t,r)=\frac{\tilde a(t)}{r^2}.
\end{equation}
Then, using suitable coordinate transformations, one can show that this solution can
be identical to the flat FRW solution. Thus, the spatially flat FRW solution is one
of the
uncharged  subclasses of our general solution (\ref{bb3}) with the parameters of $c_0=c_2=0$.

\subsubsection*{The case of $c_1=c_3=0$}
For this case, $\delta$, $h$ and $h_1$ functions vanish and the metric function $b(t,r)$
will be only a time dependant function as
\begin{equation}
b(t)=\frac{\beta(t)}{c_0}\left(1+\frac{\gamma}{\beta^2(t)}\frac{c_0}{c_2} \right).
\end{equation}
Similar to previous case, using suitable coordinate transformations, one can show that this solution can
be identical to the flat FRW solution.
\subsection*{B.~Uncharged Solution for $N=1$}\label{unn1}

\subsubsection*{1.~Uncharged solution for $\nu_0 (r)$}
One can find that both the functions $h(r)$ and $F_{01}$ in (\ref{lrvs}) and (\ref{mast}), respectively,  vanish for
\begin{equation}
\nu_0 (r)=\frac{c_5}{1+kr^2}.
\end{equation}
Thus, we have
\begin{eqnarray}
&&a(t,r)=\frac{1}{1+\frac{c_5}{a(t)} }, \\
&& b(t,r)=\frac{a(t)}{ 1+k\, r^2}\left(1+\frac{c_5}{a(t)}  \right),\label{nsmkk33}
\end{eqnarray}
Then, regarding the coordinate transformation in the section VI A 1, this solution gives also the FRW model.
\subsubsection*{2.~Uncharged solution for $ \nu_0=constant$.}\label{unchnuconst}
For this case, one can find that both the functions $h(r)$ and $F_{01}$ in (\ref{lrvs}) and (\ref{mast}), respectively,  vanish only for $k=0$ or
$\nu_0=0$. Then,
the condition for having uncharged solution for $\nu_0=constant\neq0$ is similar to
the Vaidya-Shah solution, where the uncharged case is provided only for $k=0$. For this case, we find
\begin{eqnarray}
&&a(t,r)=\frac{1}{1+\frac{\nu_0}{a(t)} }, \\
&& b(t,r)=\nu_0+a(t).
\end{eqnarray}
Accordingly, one can show that this solution also is identical to the flat FRW solution.

\section*{VII.~Apparent Horizons and Null Geodesics}

The areal distance $R$  is defined as $R=r b(t,r)$. Among the constant $R$ surfaces the null ones are called the apparent horizons. In our case there are two apparent horizons
\begin{equation}\label{null}
a^2 (b+r b^{\prime})^2-r^2 b^2 \dot{b}^2=0,
\end{equation}
where $a=q\dot b/b$. One can verify that the apparent horizons defined above reduces to
the those  given in \cite{feroni2} for the charged McVittie solution obtained
by $\nabla^{c}R\nabla_{c}R=0$ where $R$ is defined as the areal radius.  Then, there are two possibilities as
\begin{equation}\label{app1}
{\cal H}_{1}:~~q (b+r b^{\prime})-r b^2=0, ~~ \mbox{and}~~  {\cal H}_{2}: ~~q (b+r b^{\prime})+r b^2=0,
\end{equation}
for the location of the apparent horizon ${\cal H}={\cal H}_{1} \cup {\cal H}_{2}$.

According to the spacetime metric (\ref{1c}), ingoing and outgoing radial null geodesics $x^{\mu}=(t, r(t),\theta=\theta_{0}, \phi=\phi_{0})$, where $\theta_{0}$ and $\phi_{0}$ are constants, are given by
\begin{equation}\label{null1}
\frac{dr}{dt}=\pm \,q \frac{\dot b}{b^2},
\end{equation}
where $``\pm"$ signs represent the ``outgoing'' and ``ingoing'' geodesics,
respectively. These null geodesics when entered in the apparent horizon $\cal H$, they stay there. To see this, when the radial null geodesics lie in $\cal H$, by taking the derivative of $r(t)\, b(r(t),t)=c$ with respect to $t$,  we obtain
\begin{equation}\label{null2}
\frac{dr}{dt}=-\frac{r \dot b}{b+r b^{\prime}}.
\end{equation}
Eqs.(\ref{null1}) and (\ref{null2}) are consistent because the expressions in the right hand sides of these equations are equal due to the nullity condition (\ref{null}) or (\ref{app1}) of the apparent horizon $\cal H$.

To study the causal and global structures of the spacetime we have to maximally extend the existing coordinates
$\{ -\infty <t<\infty, ~ r \ge 0, ~~0 <\phi< 2 \pi,~~ 0< \theta <\pi \}$ to a coordinate system where the areal distance $R$ is one of the coordinates as done in \cite{kaloper, lake, feroni}. We postpone  a detailed
 study of this case as our future work. However,  just to give an idea how the the radial null geodesics (NG) behave we plot them  in Figure 1. In the same figure we give also apparent horizons ${\cal H}_{1}$, ${\cal H}_{2}$ and singular surface $\Sigma_{2}$ of this uncharged solution
given by
\begin{eqnarray}
&&NG:~~\frac{dr}{dt}=-\frac{r \dot b(t,r)}{b(t,r)+r b^{\prime}(t,r)},\label{nge}\\
&&{\cal H}:~~a^2(t,r)\left(b(t,r)+r b^{\prime}(t,r)\right)^2-r^2 b^2(t,r) \dot b^2(t,r)=0,\label{ahsl}\\
&&\Sigma_2:~~M^2(1+kr^2)-4a^{2}(t)(\mu+r^2)=0,\label{sigmal}
 \end{eqnarray}
respectively, corresponding to the metric functions
 \begin{eqnarray}
&&a(t,r)=\frac{1-\frac{M^2}{4a^2(t)}\, \frac{1+kr^2}{\mu+r^2} }{\left(1+\frac{M}{2a(t)}\, \sqrt{\frac{1+k\,r^2}{\mu+r^2}}\right)^2},\\
&&b(t,r)= \frac{a(t)}{1+k r^2}\,\left(1+\frac{M}{2a(t)}\, \sqrt{\frac{1+k\,r^2}{\mu+r^2}}\right)^2,
 \end{eqnarray}
for typical values of  $M, \mu$ and $k$ parameters in a de Sitter background as in the terminology
of FRW models.
\begin{figure}[]
\begin{center}
\includegraphics[scale=0.8]{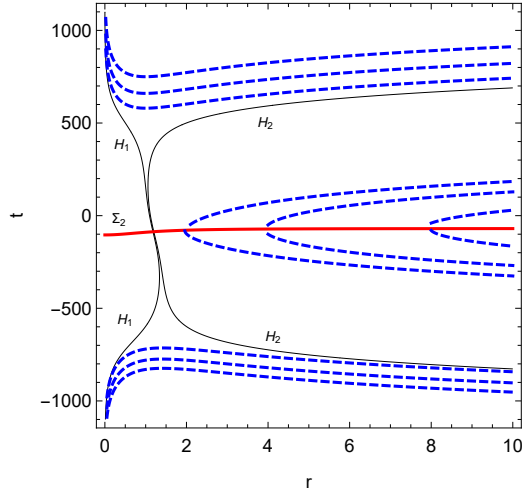}
\caption{Null geodesics (dashed blue curves), singular surface $\Sigma_2$
(thick red curve) and apparent horizons ${\cal H}_{1}$ and ${\cal H}_{2}$ (thin black curves) in the de Sitter background for the uncharged
 $N=2$ solution with $M=1, \,k=1,\, \mu=2$ and $ a(t)=e^{0.01t}$ . }
\end{center}
\end{figure}
\section*{VIII.~Conclusion}
We have found two classes of solutions of Einstein-Maxwell-Perfect Fluid field equations with a cosmological constant in a spherically symmetric spacetime. In particular the first class corresponding to the $N=2$ case contains six parameters four of which are essential generalizes the Vaidya-Shah solution. The uncharged version of our solution generalizes the McVittie solution. We showed that there are some, depending on sign of the parameters,  timelike and spacelike surfaces where the spacetime becomes singular. We then investigated some special limits of our solutions in both classes. The  list of our new solutions is given in Tables 1 and 2.

\begin{center}
\begin{table}[!ht]
\centering
\begin{tabular}{|c|c|c|c|} 
\hline\hline
$N$ &Class&Parameters &Solution
\\ [0.5ex]
\hline 
&I&$c_{0}, c_{1}, c_{2}, c_{3}\neq0$ &$b(t,r)= \frac{\beta(t)}{c_{0}+c_{1}\,r^2}\,\left(1+\frac{\delta}{\beta(t)}\, \sqrt{\frac{c_{0}+c_{1}\,r^2}{c_{2}+c_{3}\,r^2}}+\frac{\gamma}{\beta^2(t)}\, \frac{c_{0}+c_{1}\,r^2}{c_{2}+c_{3}\,r^2} \right)$~\\[2ex]
\raisebox{1.5ex}{$N=2$}
&II&$c_1=0$ or $c_3=0$& $b(t,r)=\frac{\beta(t)}{c_{0}}\,\left(1+\frac{\delta}{\beta(t)}\, \sqrt{\frac{c_0}{c_{2}+c_{3}\,r^2}}+\frac{\gamma}{\beta^2(t)}\, \frac{c_{0}}{c_2 +c_{3}\,r^2} \right)$\\[2ex]
\raisebox{1.5ex}{}
&III&$c_4=0$ & $b(t,r)= \frac{\beta(t)}{c_{0}+c_{1}r^2}+\frac{\gamma}{\beta(t)}\frac{1}{c_{2}+c_{3}\, r^2}$\\[2ex]
&IV&$c_{0}=0, c_{2} \ne 0$, or $c_{2}=0, c_{0} \ne 0$
&\mbox{Vaidya-Shah Solution}  \\[2ex]
\raisebox{1.5ex}{}
&V&$\gamma=0$  & $b(t,r)=\alpha_0(r)+ \frac{\beta(t)}{c_{0}+c_{1}r^2}$~~(identical to $N=1$)\\[2ex]
\hline
$N=1$&& $\nu_0(r)=\mbox{arbitrary},~~b_0, b_1\neq0$&$ b(t,r)=\nu_{0}(r)+\frac{\beta(t)}{ b_0+b_1 r^2}$\\
\hline\hline 
\end{tabular}
\label{tab:PPer1}
\caption{List of charged solutions and their special limits}
\end{table}
\end{center}
\vspace{2cm}

\begin{center}
\begin{table}[!ht]
\centering
\begin{tabular}{|c|c|c|c|} 
\hline\hline 
$N$& Class &Parameters &Solution
\\ [0.5ex]
\hline 
\raisebox{1.5ex}{}
&I&$\delta^2=4 \gamma$ $\&~c_{0}, c_{1}, c_{2}, c_{3}\neq0$& $b(t,r)=\left(\frac{\delta}{2\, \sqrt{\beta(t)}}\, \frac{1}{\sqrt{c_{2}+c_{3}\,r^2}}+\frac{\sqrt{\beta(t)}}{\sqrt{c_{0}+c_{1}\,r^2}} \right)^2$\\[2ex]

&II&$\delta^2=4 \gamma$, $c_{0}=0$ or $c_{2}=0$
&Generalized McVittie to non-flat Background ($k\neq0$)\\[1ex]
&&&~~(uncharged Vaidya-Shah solution)\\[2ex]\raisebox{1.5ex}{$N=2$}
&III&$\delta^2=4 \gamma$, $c_{1}=0$ or $c_{3}=0$& $b(t,r)= \left(\frac{\delta}{2\, \sqrt{~\beta(t)}}\, \frac{1}{\sqrt{c_2+c_{3}r^2}}+\frac{\sqrt{\beta(t)}}{\sqrt{c_{0}}} \right)^2$\\[2ex]
&IV&$\delta^2=4 \gamma$, $c_{1}=c_{2}=0$ or $c_{0}=c_{3}=0$&McVittie Solution\\[2ex]

\raisebox{1.5ex}{}
&V&$c_{0}c_3=c_1c_{2}$ & FRW Solution\\[2ex]
\raisebox{1.5ex}{}
&VI&$c_{0}=c_{2}=0$ & FRW Solution\\[2ex]
\raisebox{1.5ex}{}
&VII& $c_{1}=c_{3}=0$& FRW Solution\\[2ex]
\hline
&I&$\nu_0 (r)=\frac{c_5}{1+kr^2}$
&FRW Solution\\[2ex]
\raisebox{1.5ex}{$N=1$}
&II&$\nu_0=\mbox{constant}$
&FRW Solution\\[1ex]
\hline\hline 
\end{tabular}
\label{tab:PPer2}
\caption{List of uncharged solutions and their special limits}
\end{table}
\end{center}

\noindent
Among all the solutions we found in this work there are  new charged and uncharged solutions of the Einstein-Maxwell-Perfect Fluid equations with cosmological constant. For the uncharged case the solution corresponding to the $N=2$ class is a model of a universe which is a mixture of  two different FRW universes with different spatial curvatures. We will study in particular this solution in more detail in a forthcoming publication.

\vspace{0.3cm}
\noindent
Mathematical and Physical properties of our solutions can be summarized as follows:\\

\noindent
We proved three theorems: \\

\begin{description}
\item[(i)] The first theorem is on the reduction of the Einstein field equations into a single ordinary nonlinear differential equations.\\
\item[(ii)] The second theorem is on the two classes of solutions corresponding to $N=1$ and $N=2$.\\
\item[(iii)] The third theorem is on the regularity of spacetime when the radial coordinate $r$ goes to zero and to infinity.\\
\end{description}
\vspace{0.3cm}
\noindent
Other properties are the following:\\
\begin{description}
\item[(i)] Our solution corresponding to $N=2$ of Section IV when the two spatial curvatures are equal, i.e., $k_{1}=k_{2}$, reduces to FRW metric if the total charge in the universe vanishes.\\

\item[(ii)] There exits some spacelike surfaces where the pressure of the fluid diverges  but the mass density of the fluid distribution remains finite. Such spacelike surfaces are known as (sudden) cosmological singularities.\\
\item[(iii)] Null geodesics crossing the above spacelike surfaces remain in these surfaces.\\
\item[(iv)] If $\mu=0$ in the first representation in Section IV, we obtain the Vaidya-Shah metric. When $\mu=0$ and spatial curvature $k=0$ and the scale factor $a(t)=1$ we obtain the Reissner-Nordstr{\" o}m metric in isotropic coordinates. When $\mu=k=0$ and $a(t)=e^{\sqrt{\frac{\Lambda}{3}}\,t}$ then we obtain Schwarzschild-Reissner-Nordstr{\" o}m-de Sitter metric with cosmological constant $\Lambda$.
When the charge parameter $Q$ vanishes we obtain a generalization of McVittie metric. If the charge parameter vanishes and $\mu=k=0$ we get the McVittie solution. Furthermore if $a(t)=1$ we obtain the Schwarzschild metric in isotropic coordinates.\\
\item[(v)] In particular for the uncharged case our solution can be considered as a nonlinear superposition of two different FRW metrics with different scale factors and different spatial curvatures. Due to this effect in our model our universe may start with a FRW universe with spatial curvature $k_{2}$  and ends up with a FRW universe with a different spatial curvature $k_{1}$, so that $k_{1} k_{2} \le 0$. This means that the universe may undergo a change of topology.
\end{description}
\textit{Note added in the proof.-Recently, we became aware of a paper by
Mashhoon and Partovi \cite{bahram}, focusing on the gravitational collapse
of charged fluid spheres. Although the problem studied by Mashhoon and Partovi
differs from ours, they present an exact solution of Einstein field equations
for inhomogeneous charged fluid distribution which corresponds our $N=2$
solution, with zero cosmological constant, in Theorem 2.}

\section*{Appendix A: Kustaanheimo-Qvist approach for the charged case}\label{apen}
Following the Kustaanheimo-Qvist \cite{lie} approach (see also \cite{bolejku,
exact}), one can use the change of variables
 $L=b^{-1}$ and $x=r^2$ to transform the equation (\ref{denk3}) for the uncharged case ($h=0$) to the following ordinary differential equation
\begin{equation}\label{qvist}
\frac{4x}{L^2}L_{xx}+\frac{h_1}{r}=0,
\end{equation}
where using the identification $F(x)=-\frac{h_1}{4xr}$\,\, (\ref{qvist}) can be written in
the form of
\begin{equation}\label{bfg}
L_{xx}=F(x)L^2.
\end{equation}
For the case where the charge is also included we obtain
\begin{equation}\label{emtcase}
L_{xx}=F_1(x)L^2+F_2 (x)L^3,
\end{equation}
where $F_1(x)=-\frac{h_1}{4xr}$ and $F_2(x)=-\frac{h^2}{2x}$.

For the case where charge is zero,  as represented in \cite{exact}, there are three different approaches to finding solutions for (\ref{bfg}).
The first approach is based on an \textit{ad hoc} ansatz for the function $F(x)$
\cite{Sri}. The second approach is based on the answering to the question of ``for which functions $F(x)$ the equation
(\ref{bfg}) admits one (or two) Lie point symmetries or Noether symmetries?'' \cite{lie}. The third approach introduced by Wyman \cite{wyman} is based on the solutions of (\ref{bfg}) which have the Painleve property. All known solutions belong to
this class. For the case of $F=0$, the solution to (\ref{bfg}) is $\frac{1}{b}=L=A(t)r^2+B(t)$. Some other subclasses with $ F=(ax^2 +2bx+c)^{-\frac{5}{2}}$ where $a, b$ and $c$ are real constants are given in the following \cite{exact}.

\begin{description}
\item[(i)]  McVittie solution: $F(x)=\left(x(x+4R^2)\right)^{-\frac{5}{2}},~A=0$.
\item[(ii)]  $\rho=\rho(t)$ solution: $F(x)=\left( 2bx \right)^{-\frac{5}{2}},~b\neq 0$  and $6A=b(3e^{2f}-\kappa_0\mu).$
\item[(iii)]  $\rho=\rho(t)$ solution: $F(x)=0$ and $12AB= 3e^{2f}-\kappa_0\mu$.
\item[(iv)]  $p=p(\rho) $,~$\rho=\rho(t)$ solution: $F(x)=0$ and $B=\epsilon A$ where
$\epsilon=0,\pm1$.
\item[(v)]  $p=p(\rho)$,~$\rho=\rho(t,r)$: $F(x)= 1$ and $A=const,~B=t$ and $e^{-2f}=-4At$. \end{description}

\vspace{0.5cm}
\noindent
Our uncharged solution given in (\ref{bb23}) as

\begin{equation}
b(t,r)=\left(\frac{\delta}{2\, \sqrt{\beta(t)}}\, \frac{1}{\sqrt{c_{2}+c_{3}\,r^2}}+\frac{\sqrt{\beta(t)}}{\sqrt{c_{0}+c_{1}\,r^2}} \right)^2,
\end{equation}

gives a new solution to the above equation (\ref{bfg}) where the function $L$ is given by
\begin{equation}
L=\left(\frac{\delta}{2\, \sqrt{\beta(t)}}\, \frac{1}{\sqrt{c_{2}+c_{3}\,x}}+\frac{\sqrt{\beta(t)}}{\sqrt{c_{0}+c_{1}\,x}} \right)^{-2},
\end{equation}
then the function $F(x)$ is found as
\begin{equation}
F(x)=-\frac{ 3 \delta}{4}\, \frac{(c_{1}\, c_{2}-c_{0}\, c_{3})^2}{(c_{0}+c_{1}\,x)^{5/2}\,(c_{2}+c_{3}\,x)^{5/2}}.
\end{equation}

For the charged case we will investigate all possible new
exact solutions and the properties of the equation (\ref{emtcase}) in a later communication.


\section*{Appendix B: The case of  $\mu=\frac{c_{2}}{c_{3}}$}\label{dafne00}

 For this case, $S(t,r), \,X(t,r)$ and $Y(t,r)$ functions are given
by the following forms.
\begin{eqnarray}
S(t,r)&=&\frac{3(2M^2-Q^2)(1-k\mu)r^2}{(1+kr^2)^3 (\mu+r^2)^3}-\frac{12a^2(t)k}{(1+kr^2)^4 }\nonumber\\
&&-\frac{6Ma(t)}{(1+kr^2)^{\frac{7}{2}} (\mu+r^2)^{\frac{5}{2}}}\left(\mu+3k\mu^2+k(3+\mu
k)r^4+8k\mu r^2\right)\nonumber\\
&&-\frac{3M^{2}(1+k\mu +2kr^2)}{(1+kr^2)^{3} (\mu+r^2)^{3}}\left(2\mu+k\mu r^2+r^{2}\right)-\frac{3(M^{2}-Q^2)}{(1+kr^2)^{2} (\mu+r^2)^{2}}\left(1+k\mu
\right)\nonumber\\
&&-\frac{3M(M^{2}-Q^2)}{2a(t)(1+kr^2)^{\frac{5}{2}} (\mu+r^2)^{\frac{7}{2}}}\left(3\mu+8k\mu
r^2+k(1+3\mu
k)r^4+k\mu^2\right)\nonumber\\
&&-\frac{3(M^2-Q^2)^2 \mu}{4a^2(t)(\mu+r^2)^4},
\end{eqnarray}
and
\begin{eqnarray}
X(t,r)&=&\frac{1+\frac{M}{a(t)}\, \sqrt{\frac{1+k\,r^2}{\mu+r^2}}+\frac{M^2-Q^2}{4a^2(t)}\frac{1+kr^2}{\mu+r^2}}{1-\frac{M^2-Q^2}{4a^2(t)}\, \frac{1+kr^2}{\mu+r^2}},\\
Y(t,r)&=&-\frac{4ka^2(t)}{(1+kr^2)^4}-\frac{8kMa(t)}{(1+kr^2)^{\frac{7}{2}}(\mu+r^2)^{\frac{1}{2}}}\nonumber\\
&&-\frac{4M^2k(\mu+2\mu r^2 +r^4)-(M^2-Q^2)\left[\mu(1-2k\mu)+2k(1-\mu)r^2
+k(k\mu-1)r^4  \right]}{(1+kr^2)^{3}(\mu+r^2)^{3}}\nonumber\\
&&+\frac{1}{a(t)}\left[\frac{2M(M^2-Q^2)(1-k\mu)\left[\mu- kr^4\right]}{(1+kr^2)^{\frac{5}{2}}(\mu+r^2)^{\frac{7}{2}}}\right]\nonumber\\
&&+\frac{1}{a^2(t)}\Big[\frac{(M^2-Q^2)^{2}\left[\mu(2-k\mu)+2k\mu r^2+k(2k\mu-1)r^4\right]+M^2(M^2-Q^2)\left[4\mu+8k(\mu+1)+4k(k\mu-2)r^4\right]}{4(1+kr^2)^{2}(\mu+r^2)^{4}}\Big]\nonumber\\
&&+\frac{1}{a^3(t)}\left[\frac{M(M^2-Q^2)^2\mu(1+kr^2)^{\frac{1}{2}}}{2(\mu+r^2)^{\frac{9}{2}}}\right]+\frac{1}{a^4(t)}\left[\frac{\mu(M^2-Q^2)^3
(1+kr^2)}{16(\mu+r^2)^{5}}\right].
\end{eqnarray}

At the spatial origin, i.e $r\rightarrow 0$, the  behavior  of the functions $a(t,r), \, b(t,r), \,\rho(t,r), \,\sigma(t,r)$ and $p(t,r)$ are given by\begin{eqnarray}
&&a(t,r)\rightarrow \frac{1-\frac{M^2-Q^2}{4a^2(t)\mu}}{\left(1+\frac{M}{a(t)\sqrt{\mu}}+\frac{M^2-Q^2}{4a^2(t)\mu} \right)},\label{agoli}\\
&&b(t,r)\rightarrow  a(t)\,\left(1+\frac{M}{a(t)\sqrt{\mu}}+\frac{M^2-Q^2}{4a^2(t)\mu}\, \right),\\
&&4 \pi \sigma(t,r)\rightarrow \frac{3 |Q|\left(1-\mu k\right)a^3(t)}{\mu^{\frac{3}{2}}\,\left(a^2(t)+\frac{Ma(t)}{\sqrt{\mu}}+\frac{M^2-Q^2}{4\mu}\, \right)^3},\label{s11}\\
&&8\pi \rho(t,r)\rightarrow3\frac{{\dot a}^2(t)}{a^2(t)}+3a^2(t)S_0(t)-\Lambda,\label{rho11}\\
&&8\pi p(t,r)\rightarrow-3\frac{\dot a^2(t)}{a^2(t)}+2\left(\frac{\dot a^2(t)}{a^2(t)}-\frac{\ddot a(t)}{a(t)}\right)X_0(t)\nonumber\\
&&~~~~~~~~~~~~~~~~+\frac{Y_0(t)}{a^4(t)\left( 1-\frac{M^2-Q^2}{4\mu a^2(t)} \right)\left(1+\frac{M}{a(t)\sqrt{\mu}}+\frac{M^2-Q^2}{4a^2(t)\mu}\, \right)^4}+\Lambda,
\end{eqnarray}
where $\mu>0$ and $S_{0}(t), ~X_{0}(t)$ and $Y_{0}(t)$ read as\begin{eqnarray}
S_{0}(t)&=&64\Big[\frac{16ka^{4}(t)\mu^4+8Ma^3(t)\mu^{\frac{5}{2}}(1+3k\mu)+4(3M^2-Q^2)(1+k\mu)a^{2}(t)\mu^2
}{ \,\left(4a^{2}(t)\mu+4Ma(t)\sqrt\mu+M^2-Q^2 \right)^4},\nonumber\\
&&~~~~~~+\frac{2a(t)\mu^{\frac{3}{2}}M(M^2-Q^2)(3+k\mu)+\mu(M^2-Q^2)^2
}{\left(4a^{2}(t)\mu+4Ma(t)\sqrt\mu+M^2-Q^2 \right)^4}\Big],\\
X_0(t)&=&\frac{1+\frac{M}{a(t)\sqrt{\mu}}+\frac{M^2-Q^2}{4a^2(t)\mu}}{1-\frac{M^2-Q^2}{4\mu
a^2(t)}},\\
Y_{0}(t)&=&-4ka^2(t)-\frac{8kMa(t)}{\sqrt\mu}-\frac{4M^2k-(M^2-Q^2)(1-2k\mu)}{\mu^{2}}+\frac{1}{a(t)}\left[\frac{2M(M^2-Q^2)(1-k\mu)\mu}{\mu^{\frac{7}{2}}}\right]\nonumber\\
&&+\frac{1}{a^2(t)}\left[\frac{(M^2-Q^2)^{2}(2-k\mu)+M^2(M^2-Q^2)\left[4\mu+8k(\mu+1)\right]}{4\mu^{4}}\right]\nonumber\\
&&+\frac{1}{a^3(t)}\left[\frac{M(M^2-Q^2)^2}{2\mu^{\frac{7}{2}}}\right]+\frac{1}{a^4(t)}\left[\frac{(M^2-Q^2)^3}{16\mu^{4}}\right].
\end{eqnarray}
Then, all the functions $a(t,r), \, b(t,r),\, \rho(t,r),\, \sigma(t,r)$ and $p(t,r)$ remain finite at the spatial origin.

At the spatial infinity, i.e $r\rightarrow \infty$, assuming $\mu>0$ and $k>0$ the behavior  of the functions $a(t,r),~b(t,r)$, $\rho(t,r),~\sigma(t,r)$
and $p(t,r)$ are
given by
\begin{eqnarray}
&&a(t,r)\rightarrow  \frac{1-\frac{(M^2-Q^2)k}{4a^2(t)}}{\left(1+\frac{M\sqrt
k}{a(t)}+\frac{(M^2-Q^2)k}{4a^2(t)} \right)},\\
&&b(t,r) \rightarrow 0,\\
&&4 \pi \sigma(t,r)\rightarrow
 -\frac{3|Q|(1-\mu k)k^{\frac{3}{2}}}{a^3(t)\left(1+\frac{M\sqrt k}{a(t)}+\frac{M^2-Q^2}{4a^2(t)}k
 \right)^3},\\
&&8\pi \rho(t,r)\rightarrow 3\frac{{\dot a}^2(t)}{a^2(t)}+3a^2(t)S_1(t)-\Lambda,\\
&&8\pi p(t,r)\rightarrow-3\frac{\dot a^2(t)}{a^2(t)}+2\left(\frac{\dot a^2(t)}{a^2(t)}-\frac{\ddot a(t)}{a(t)}\right)X_1(t)\\
&&~~~~~~~~~~~~~~~~+\frac{Y_1(t)}{\left( 1-\frac{k(M^2-Q^2)}{4 a^2(t)} \right)\left(1+\frac{M\sqrt
k}{a(t)}+\frac{(M^2-Q^2)k}{4a^2(t)} \right)^4}+\Lambda,
\end{eqnarray}
where $k\geq 0$ and $S_1(t),\,X_1(t)$ and $Y_1(t)$ are given by
\begin{eqnarray}
S_1(t)&=&64\Big[\frac{16ka^{4}(t)+8Ma^3(t)k^{\frac{3}{2}}(3+k\mu)+4(3M^2-Q^2)(1+k\mu)a^{2}(t)k^2
}{ \,\left(4a^{2}(t)+4Ma(t)\sqrt k+(M^2-Q^2)k \right)^4},\\
&&~~~~~+\frac{2a(t)M k^{\frac{5}{2}}(M^2-Q^2)(1+3k\mu)+(M^2-Q^2)^2 k^4\mu
}{\left(4a^{2}(t)+4Ma(t)\sqrt k+(M^2-Q^2)k\right)^4}\Big],\\
X_1(t)&=&\frac{1+\frac{M}{a(t)}\, \sqrt{k}+\frac{k(M^2-Q^2)}{4a^2(t)}}{1-\frac{k(M^2-Q^2)}{4a^2(t)}\, },\\
Y_1(t)&=&-\frac{4a^2(t)}{k^3}-\frac{8Ma(t)}{k^{\frac{5}{2}}}-\frac{4M^2 +(M^2-Q^2)(k\mu-1)}{k^2}\nonumber\\
&&+\frac{1}{a(t)}\left[\frac{2M(M^2-Q^2)(k\mu-1)}{k^{\frac{3}{2}}}\right]+
\frac{1}{a^2(t)}\Big[\frac{(M^2-Q^2)^{2}(2k\mu-1)+4M^{2}(M^2-Q^2)(k\mu-2)}{4k}\Big]\nonumber\\
&&+\frac{1}{a^3(t)}\left[\frac{M(M^2-Q^2) k^{\frac{1}{2}}\mu }{2}\right]
+\frac{1}{a^4(t)}\left[\frac{(M^2-Q^2)^3 k\mu}{16}\right].
\end{eqnarray}
Then, all functions $a(t,r), \, b(t,r), \,\rho(t,r),\, \sigma(t,r)$ and $p(t,r)$ remain regular at the asymptotic region. Also, it seen that regarding
the above forms of $\rho(t,r), \sigma(t,r)$ and $p(t,r)$ functions, the behavior of this solution at the spatial infinity is different than FRW solution.

\section*{Appendix C: The case of $k_2=\frac{c_{3}}{c_{2}}$}\label{dafne11}
For this case, $S(t,r), \,X(t,r)$ and $Y(t,r)$ functions are given
by the following forms.
\begin{eqnarray}
S(t,r)&=&\frac{3(2M^2-Q^2)(k_{2}-k_{1})^{2}r^2}{(1+k_{1}r^2)^3 (1+k_2 r^2)^3}-\frac{12a^2(t)k_1}{(1+k_1r^2)^4 }\nonumber\\
&&-\frac{6Ma(t)}{(1+k_1r^2)^{\frac{7}{2}} (1+k_2r^2)^{\frac{5}{2}}}\left(k_2+3k_1+k_1
k_2(k_{1}+3k_2)r^4+8k_1 k_2r^2\right)\nonumber\\
&&-\frac{3M^{2}(k_1+k_2 +2k_1 k_2r^2)}{(1+k_1r^2)^{3} (1+k_2r^2)^{3}}\left(2+(k_1+k_{2})r^{2}\right)-\frac{3(M^{2}-Q^2)}{(1+k_1r^2)^{2} (1+k_2r^2)^{2}}\left(k_1+k_2
\right)\nonumber\\
&&-\frac{3M(M^{2}-Q^2)}{2a(t)(1+k_1r^2)^{\frac{5}{2}} (1+k_2r^2)^{\frac{7}{2}}}\left(k_{1}+3k_2+8k_1
k_2 r^2+k_1 k_2(k_2+3k_1)r^4\right)\nonumber\\
&&-\frac{3(M^2-Q^2)^2 k_2}{4a^2(t)(1+k_2r^2)^4},\\
X(t,r)&=&\frac{1+\frac{M}{a(t)}\, \sqrt{\frac{1+k_1\,r^2}{1+k_2r^2}}+\frac{M^2-Q^2}{4a^2(t)}\, \frac{1+k_1r^2}{1+k_2r^2}}{1-\frac{M^2-Q^2}{4a^2(t)}\, \frac{1+k_1r^2}{1+k_2r^2}},\\
Y(t,r)&=&-\frac{4k_1a^2(t)}{(1+k_1r^2)^4}-\frac{8k_{1}Ma(t)}{(1+k_{1}r^2)^{\frac{7}{2}}(1+k_2r^2)^{\frac{1}{2}}}\nonumber\\
&&-\frac{4M^2k_1(1+2k_2r^2 +k_2^2r^4)-(M^2-Q^2)\left[(k_{2}-2k_1)+2k_{1}k_2(k_2-1)r^2
+k_1k_2(k_1-k_2)r^4  \right]}{(1+k_1r^2)^{3}(1+k_2r^2)^{3}}\nonumber\\
&&+\frac{1}{a(t)}\left[\frac{2M(M^2-Q^2)(k_2-k_1)\left[1- k_1k_2r^4\right]}{(1+kr^2)^{\frac{5}{2}}(\mu+r^2)^{\frac{7}{2}}}\right]\nonumber\\
&&+\frac{1}{a^2(t)}\Big[\frac{(M^2-Q^2)^{2}\left[2k_2-k_1+2k_1k_2 r^2+k_1
k_2(2k_1-k_2)r^4\right]}{4(1+k_1r^2)^{2}(1+k_2r^2)^{4}}\nonumber\\
&&~~~~~~~~~~~+\frac{M^2(M^2-Q^2)\left[4k_2+8k_1k_2(1+k_2)+4k_1k_2(k_1-2k_2)r^4\right]}{4(1+k_1r^2)^{2}(1+k_2r^2)^{4}}\Big]\nonumber\\
&&+\frac{1}{a^3(t)}\left[\frac{M(M^2-Q^2)^2k_2(1+k_1r^2)^{\frac{1}{2}}}{2(1+k_2r^2)^{\frac{9}{2}}}\right]+\frac{1}{a^4(t)}\left[\frac{k_{2}(M^2-Q^2)^3
(1+k_1r^2)}{16(1+k_2r^2)^{5}}\right],
\end{eqnarray}

 At the spatial origin, i.e $r\rightarrow 0$, the behavior  of the functions $a(t,r), \, b(t,r),\, \rho(t,r), \,\sigma(t,r)$ and $p(t,r)$  are given by
\begin{eqnarray}
&&a(t,r)\rightarrow\frac{1-\frac{M^2-Q^2}{4a^2(t)}}{\left(1+\frac{M}{a(t)}+\frac{M^2-Q^2}{4a^2(t)} \right)},\label{masi}\\
&&b(t,r)\rightarrow a(t)\,\left(1+\frac{M}{a(t)}+\frac{M^2-Q^2}{4a^2(t)}\, \right),\\
&&4 \pi \sigma(t,r)\rightarrow \frac{192|Q|\left(k_2- k_1\right)a^3(t)}{\left(4a^2(t)+4Ma(t)
+M^2-Q^2 \right)^3},\label{s11s}\\
&&8\pi \rho(t,r)\rightarrow3\frac{{\dot a}^2(t)}{a^2(t)}+a^2(t)S_0(t)-\Lambda,\label{rho11t}\\
&&8\pi p(t,r)\rightarrow-3\frac{\dot a^2(t)}{a^2(t)}+2\left(\frac{\dot a^2(t)}{a^2(t)}-\frac{\ddot a(t)}{a(t)}\right)X_0(t)\nonumber\\
&&~~~~~~~~~~~~~~~~+\frac{Y_0(t)}{a^4(t)\left( 1-\frac{M^2-Q^2}{4 a^2(t)} \right)\left(1+\frac{M}{a(t)}+\frac{M^2-Q^2}{4a^2(t)}\, \right)^4}+\Lambda\label{p11},
\end{eqnarray}
where  $S_{0}(t),~ X_0(t)$ and $Y_0(t)$ functions are
\begin{eqnarray}
S_0(t)&=&64\Big[\frac{16k_1a^{4}(t)+8Ma^{3}(t)(k_2+3k_1)+4a^{2}(t)(3M^2-Q^2)(k_1+k_2)
}{ \,\left(4a^{2}(t)+4Ma(t)+(M^2-Q^2) \right)^4}\nonumber\\
&&~~~~~~~~+\frac{2a(t)M (M^2-Q^2)(k_1+3k_2)+(M^2-Q^2)^2 k_2
}{ \,\left(4a^{2}(t)+4Ma(t)+(M^2-Q^2) \right)^4}\Big],\\
X_0(t)&=&\frac{1+\frac{M}{a(t)}+\frac{M^2-Q^2}{4a^2(t)}}{1-\frac{M^2-Q^2}{4
a^2(t)}},
\end{eqnarray}
and
\begin{eqnarray}
Y_{0}(t)&=&-4k_{1}a^2(t)-8k_1Ma(t)-4M^2k_1+(M^2-Q^2)(k_2-2k_1)
+\frac{1}{a(t)}\left[2M(M^2-Q^2)(k_2-k_1)\right]\nonumber\\
&&+\frac{1}{a^2(t)}\left[\frac{(M^2-Q^2)^{2}(2k_2-k_1)+M^2(M^2-Q^2)\left[4k_2+8k_1k_2(1+k_2)\right]}{4}\right]\nonumber\\
&&+\frac{1}{a^3(t)}\left[\frac{Mk_2(M^2-Q^2)^2}{2}\right]+\frac{1}{a^4(t)}\left[\frac{k_2(M^2-Q^2)^3}{16}\right].
\end{eqnarray}

Then, similar to the metric functions, we see that   $\sigma(t,r)$, $\rho(t,r)$ and $p(t,r)$ are regular at $r\rightarrow 0$, except  for cosmologies
with $a(t)\rightarrow 0$.

At the spatial infinity, i.e $r\to \infty$, assuming $k_{1}>0$ and $k_{2}>0$
the behavior  of the functions $a(t,r),~b(t,r)$, $\rho(t,r)$, $~\sigma(t,r)$
 and $p(t,r)$ are given by
 \begin{eqnarray}
&&a(t,r)\rightarrow  \frac{1-\frac{(M^2-Q^2)k_{1}}{4a^2(t)k_2} }{\left(1+\frac{M\sqrt
k_1}{a(t)\sqrt k_2}+\frac{(M^2-Q^2)k_1}{4a^2(t)k_2} \right)} \label{newaswru},\\
&&b(t,r) \rightarrow 0,\\
&&4 \pi \sigma(t,r)\rightarrow
 \frac{192|Q|\left(k_2- k_1\right)(k_1 k_2)^{\frac{3}{2}}a^3(t)}{\left(4a^2(t)k_2+4Ma(t)
\sqrt{k_1 k_2}+(M^2-Q^2)k_1 \right)^3}, \\
&&8\pi \rho(t,r)\rightarrow 3\frac{{\dot a}^2(t)}{a^2(t)}+3a^2(t)S_1(t)-\Lambda,\\
&&8\pi p(t,r)\rightarrow-3\frac{\dot a^2(t)}{a^2(t)}+2\left(\frac{\dot a^2(t)}{a^2(t)}-\frac{\ddot a(t)}{a(t)}\right)X_1(t)\nonumber\\
&&~~~~~~~~~~~~~~~+\frac{Y_1(t)}{\left( 1-\frac{(M^2-Q^2)k_{1}}{4 a^2(t)k_2} \right)\left(1+\frac{M\sqrt
k_1}{a(t)\sqrt k_2}+\frac{(M^2-Q^2)k_1}{4a^2(t)k_2} \right)^4}+\Lambda,
\end{eqnarray}
where  $S_{1}(t),~ X_1(t)$ and $Y_1(t)$ functions read as
\begin{eqnarray}
S_1(t)&=&64\Big[\frac{16k_1k_2^4a^{4}(t)+8Ma^{3}(t){k_1}^{\frac{3}{2}}{k_2}^{\frac{5}{2}}(k_1+3k_2)
+4a^{2}(t)(3M^2-Q^2)k_1^2 k_2^2(k_1+k_2)
}{ \,\left(4a^{2}(t)k_2+4Ma(t)\sqrt{k_1 k_2}+(M^2-Q^2)k_1 \right)^4}\nonumber\\
&&~~~~~+\frac{2a(t)M k_1^{\frac{5}{2}}k_2^{\frac{3}{2}}(M^2-Q^2)(k_2+3k_1)+(M^2-Q^2)^2 k_2k_1^4}{ \,\left(4a^{2}(t)k_2+4Ma(t)\sqrt{k_1 k_2}+(M^2-Q^2)k_1 \right)^4}\Big],\\
X_1(t)&=&\frac{1+\frac{M}{a(t)}\, \sqrt{\frac{k_1}{k_2}}+\frac{(M^2-Q^2)k_{1}}{4a^2(t)k_2}}{1-\frac{(M^2-Q^2)k_1}{4a^2(t)k_2}\, },\end{eqnarray}
and
\begin{eqnarray}
Y_1(t)&=&-\frac{4a^2(t)}{k_1^3}-\frac{8Ma(t)}{k_1^{\frac{5}{2}}}-\frac{4M^2k_2 +(M^2-Q^2)(k_1-k_2)}{k_1^2k_2^2}\nonumber\\
&&+\frac{1}{a(t)}\left[\frac{2M(M^2-Q^2)(k_1-k_2)}{k_1^{\frac{3}{2}}k_2^{\frac{5}{2}}}\right]\nonumber\\
&&+\frac{1}{a^2(t)}\Big[\frac{(M^2-Q^2)^{2}(2k_1-k_2)+4M^{2}(M^2-Q^2)(k_1-2k_2)}{4k_1k_2^3}\Big]\nonumber\\
&&+\frac{1}{a^3(t)}\left[\frac{M(M^2-Q^2) k^{\frac{1}{2}}}{2k_2^{\frac{7}{2}}}\right]+\frac{1}{a^4(t)}\left[\frac{(M^2-Q^2)^3 k_1}{16k_2^4}\right].
\end{eqnarray}
Then, all functions $a(t,r), \, b(t,r), \,\rho(t,r), \sigma(t,r)$ and $p(t,r)$ remain regular at the asymptotic region. Also, one realizes that regarding
the above forms of $\rho(t,r), \sigma(t,r)$ and $p(t,r)$ functions, the behavior of this solution at the spatial infinity is different than FRW solution.

\section*{Appendix D: The case of either $c_{1}=0$ or $c_{3}=0$}\label{bito}
For this case, we have
\begin{eqnarray}
S(t,r)&=&\frac{6Mka(t)}{(1+kr^2)^{\frac{5}{2}}}+
\frac{3k(3M^2-Q^2)}{(1+kr^2)^3}+\frac{9M(M^2-Q^2)k}{2a(t)(1+kr^2)^{\frac{7}{2}}}
+\frac{3k(M^2-Q^2)^2}{4a^{2}(t)(1+kr^2)^4},\\
X(t,r)&=&\frac{1+\frac{M}{a(t)}\, \frac{1}{\sqrt{1+kr^2}}+\frac{M^2-Q^2}{4a^2(t)}\, \frac{1}{1+kr^2}}{1-\frac{M^2-Q^2}{4a^2(t)}\, \frac{1}{1+kr^2}},\\
Y(t,r)&=&\frac{k(M^2-Q^2)}{(1+kr^2)^{3}}+\frac{1}{a(t)}\left[\frac{2M(M^2-Q^2)k}{(1+kr^2)^{\frac{7}{2}}}\right]+\frac{1}{a^2(t)}\left[\frac{(M^2-Q^2)^{2}k+2M^2(M^2-Q^2)k}{(1+kr^2)^4}\right]\nonumber\\
&&+\frac{1}{a^3(t)}\left[\frac{M(M^2-Q^2)^2k}{2(1+kr^2)^{\frac{9}{2}}}\right]+\frac{1}{a^4(t)}\left[\frac{k(M^2-Q^2)^3}{16(1+kr^2)^{5}}\right].
\end{eqnarray}
\section*{Appendix E: The case  $c_{4}=0$}\label{bitol}
For this solution, we have
\begin{eqnarray}
S(t,r)&=&-\frac{3Q^2(k_{2}-k_{1})^{2}r^2}{(1+k_{1}r^2)^3 (1+k_2 r^2)^3}-\frac{12a^2(t)k_1}{(1+k_1r^2)^4 }\nonumber\\
&&+\frac{3Q^2}{(1+k_1r^2)^{2} (1+k_2r^2)^{2}}\left(k_1+k_2
\right)+\frac{3Q^4 k_2}{4a^2(t)(1+k_2r^2)^4},\\
X(t,r)&=&\frac{1-\frac{Q^2}{4a^2(t)}\, \frac{1+k_1r^2}{1+k_2r^2}}{1+\frac{Q^2}{4a^2(t)}\, \frac{1+k_1r^2}{1+k_2r^2}},\\
Y(t,r)&=&-\frac{4k_1a^2(t)}{(1+k_1r^2)^4}-\frac{Q^2\left[(k_{2}-2k_1)+2k_{1}k_2(k_2-1)r^2
+k_1k_2(k_1-k_2)r^4  \right]}{(1+k_1r^2)^{3}(1+k_2r^2)^{3}}\nonumber\\
&&+\frac{1}{a^2(t)}\left[\frac{Q^{4}\left[2k_2-k_1+2k_1k_2 r^2+k_1
k_2(2k_1-k_2)r^4\right]}{4(1+k_1r^2)^{2}(1+k_2r^2)^{4}}\right]-\frac{1}{a^4(t)}\left[\frac{k_{2}Q^6
(1+k_1r^2)}{16(1+k_2r^2)^{5}}\right].
\end{eqnarray}
%
\section*{Appendix F: Reduction to the FRW Solutions}
In Section VI A 1, we showed that when the metric function $b(t,r)$ takes the form of
\begin{equation}
b(t,r)=\frac{R(t)}{1+kr^2},
\end{equation}
the corresponding spacetime metric reduces to the FRW metric.
Here, we show that how the corresponding matter density at both  $a(t)\rightarrow
0$ and $a(t)\rightarrow \infty$ limits reduce to the matter density given
by the standard Friedman equation for an
FRW universe.
\begin{description}
\item[(i)]  For $a(t)\rightarrow 0$, regarding (\ref{kcr}), we have
\begin{equation}
b(t,r)\rightarrow\frac{M^2}{4 a(t)}\,\frac{1}{1+k_{2}\, r^2}.
\end{equation}
The corresponding matter density $\rho(t,r)$ can be read from (\ref{rhotho})
as
 \begin{equation}
8\pi\rho(t,r)\rightarrow\frac{3}{q^2}+3
\frac{ 64k_2 a^2(t)}{M^4}-\Lambda,
\end{equation}
where using $q(t)=\frac{a(t)}{\dot a(t)}$ and  $R(t)=\frac{M^2}{4a(t)}$
reduces to the following standard Friedmann equation
 \begin{equation}
8\pi\rho(t,r)\rightarrow 3\frac{\dot R(t)}{R^2(t)}+
3\frac{4k_2 }{R^2(t)}-\Lambda,
\end{equation}
describing an FRW universe with the scale factor $R(t)$ and spatial curvature $4k_2$.
\item[(i)]  For the case of $a(t)\rightarrow \infty$, from (\ref{kcr}), we
have
\begin{equation}
b(t,r)\rightarrow\frac{a(t)}{1+k_{1}r^2}.
\end{equation}
The corresponding matter density can be obtained from  (\ref{rhotho}) as
 \begin{equation}
8\pi\rho(t,r)\rightarrow\frac{3}{q^2}+
3\frac{ 4k_1}{a^2(t)}-\Lambda,
\end{equation}
where it represents the matter density of an FRW universe with the scale factor $R(t)=a(t)$
and spatial curvature $4k_1$.
 \end{description}

\end{document}